\documentclass[]{article}

\usepackage{savesym}
\usepackage{graphicx}

\usepackage{authblk}

\savesymbol{tablenum}
\usepackage{siunitx}
\newcommand\unit[3][]{\SI[#1]{#2}{#3}}
\restoresymbol{SIX}{tablenum}

\usepackage[version=3]{mhchem}
\usepackage{geometry}
\usepackage[utf8]{inputenc}       % mit fontenc: Umlaute und ß (sz) direkt eingeben
\usepackage[T1]{fontenc}          % ^...^

\usepackage{amsmath}
\usepackage{amssymb}
\usepackage{mathrsfs}             % calligraphie

\usepackage{lineno}

\usepackage{tikz}
\usetikzlibrary{plotmarks,patterns,calc,decorations.markings}

\usepackage{booktabs}

\catcode`\@=11 % @ signs are now treated as letters
\def\parenbar{\mathpalette\p@renb@r}
\def\p@renb@r#1#2{\vbox{%
\ifx#1\scriptscriptstyle \dimen@.7em\dimen@ii.2em\else
\ifx#1\scriptstyle \dimen@.8em\dimen@ii.25em\else
\dimen@1em\dimen@ii.4em\fi\fi \offinterlineskip
\ialign{\hfill##\hfill\cr
\vbox{\hrule width\dimen@ii}\cr
\noalign{\vskip-.3ex}%
\hbox to\dimen@{$\mathchar300\hfil\mathchar301$}\cr
\noalign{\vskip-.3ex}%
$#1#2$\cr}}}
\catcode`\@=12 % @ signs are no longer letters
\def\nuan{\parenbar{\nu}\kern-0.4ex}

\newcommand{\BigO}[1]{\ensuremath{\operatorname{O}\bigl(#1\bigr)}}

\newlength{\smfigwidth}
\newlength{\figwidth}
\newlength{\captwidth}
\usepackage[pdflatex]{trimclip}

\usepackage{hyperref}
\usepackage[all]{hypcap}          % reference not to captions but to Figure/table itself

\def\unitFluxNorm {\giga\electronvolt\,\centi\metre\tothe{$-2$}\,\second\tothe{$-1$}}

\title{First all-flavour Neutrino Point-like Source Search with the ANTARES Neutrino Telescope}

\date{}

\author[1]{A.~Albert}
\author[2]{M.~Andr\'e}
\author[3]{M.~Anghinolfi}
\author[4]{G.~Anton}
\author[5]{M.~Ardid}
\author[6]{J.-J.~Aubert}
\author[7]{T.~Avgitas}
\author[7]{B.~Baret}
\author[8]{J.~Barrios-Mart\'{\i}\thanks{Corresponding author}}
\author[9]{S.~Basa}
% ...

\affil[1]{\scriptsize{GRPHE - Universit\'e de Haute Alsace - Institut universitaire de technologie de Colmar, 34 rue du Grillenbreit BP 50568 - 68008 Colmar, France}}
\affil[2]{\scriptsize{Technical University of Catalonia, Laboratory of Applied Bioacoustics, Rambla Exposici\'o, 08800 Vilanova i la Geltr\'u, Barcelona, Spain}}
\affil[3]{\scriptsize{INFN - Sezione di Genova, Via Dodecaneso 33, 16146 Genova, Italy}}
\affil[4]{\scriptsize{Friedrich-Alexander-Universit\"at Erlangen-N\"urnberg, Erlangen Centre for Astroparticle Physics, Erwin-Rommel-Str. 1, 91058 Erlangen, Germany}}
\affil[5]{\scriptsize{Institut d'Investigaci\'o per a la Gesti\'o Integrada de les Zones Costaneres (IGIC) - Universitat Polit\`ecnica de Val\`encia. C/  Paranimf 1, 46730 Gandia, Spain}}
\affil[6]{\scriptsize{Aix Marseille Univ, CNRS/IN2P3, CPPM, Marseille, France}}
\affil[7]{\scriptsize{APC, Univ Paris Diderot, CNRS/IN2P3, CEA/Irfu, Obs de Paris, Sorbonne Paris Cit\'e, France}}
\affil[8]{\scriptsize{IFIC - Instituto de F\'isica Corpuscular (CSIC - Universitat de Val\`encia) c/ Catedr\'atico Jos\'e Beltr\'an, 2 E-46980 Paterna, Valencia, Spain}}
\affil[9]{\scriptsize{LAM - Laboratoire d'Astrophysique de Marseille, P\^ole de l'\'Etoile Site de Ch\^ateau-Gombert, rue Fr\'ed\'eric Joliot-Curie 38,  13388 Marseille Cedex 13, France}}
% ...

\begin{document}

\maketitle

\begin{abstract}

A search for cosmic neutrino sources using the data collected with the ANTARES neutrino telescope between early 2007 and the end of 2015 is performed. 
    For the first time, all neutrino interactions --charged and neutral current interactions of all flavors-- are considered in a search for point-like sources with the ANTARES detector. In previous analyses, only muon neutrino charged current interactions were used. This is achieved by using a novel reconstruction algorithm for shower-like events in addition to the standard muon track reconstruction.
    The shower channel contributes about 23\% of all signal events for an $E^{-2}$ energy spectrum. No significant excess over background is found. 
    The most signal-like cluster of events is located at $(\alpha,\delta) = (\unit{343.8}{\degree}, \unit{23.5}{\degree})$ with a significance of $1.9\sigma$.
    The neutrino flux sensitivity of the search is about $E^2 d\varPhi/dE = \unit{6e-9} {\unitFluxNorm}$ for declinations from $\unit{-90}{\degree}$ up to $\unit{-42}{\degree}$, and  below $\unit{e-8} {\unitFluxNorm}$ for declinations up to $\unit{5}{\degree}$.
    The directions of 106 source candidates and of 13 muon track events from the IceCube HESE sample are investigated for a possible neutrino signal and upper limits on the signal flux are determined.

\end{abstract}

\maketitle

\section{Introduction}

Different types of astrophysical objects have been proposed as production sites of high-energy neutrinos through the decay of charged pions, previously produced in the interactions of nuclei with ambient matter or radiation \cite{Sources1, Sources2, Sources3, Sources4, Sources5}.
In contrast to charged cosmic rays, neutrinos are not deflected by (inter-)galactic magnetic fields and point straight back to their production sites.
Finding sources of cosmic neutrinos would identify sources of cosmic rays, whose origin and acceleration processes are a long-standing astrophysical question.
The results of the latest search for point-like sources using the ANTARES neutrino telescope are presented in this paper. For the first time, events based on the signal
induced by electromagnetic and/or hadronic showers are included. This has been achieved by using a new reconstruction algorithm \cite{TMichael_proc,TANTRA} which allows a median pointing accuracy between \unit{2}{\degree} and \unit{3}{\degree} for electron neutrinos which interact via CC with energies in the 10$^3$ -- 10$^6$ GeV range.

This paper is structured as follows: in Sec. \ref{sec:ANTARES}, the ANTARES neutrino telescope and the event selection for this analysis are introduced. The search method is explained in Sec. \ref{sec:METHOD}, the results of the analysis shown in Sec.\ref{sec:RESULTS} and the discussion on the possible effects due to systematic uncertainties described in Sec. \ref{sec:SYST}. Finally, the conclusions are summarized in Sec. \ref{sec:CONCL}.

\section{ANTARES neutrino telescope and event selection}\label{sec:ANTARES}

The ANTARES telescope \cite{antares}, located in the Mediterranean Sea, is the largest neutrino detector in the Northern Hemisphere.
The detector comprises a three-dimensional array of 885 optical modules (OMs), each one housing a 10'' photomultiplier tube (PMT), and distributed over 12 vertical strings anchored at the sea floor at a depth of about 2400 m. 
The detection of light from upgoing charged particles is optimized with the PMTs facing \unit{45}{\degree} downward. 

Simulations are used to evaluate the performance of the detector. Atmospheric muons and neutrinos are simulated with the MUPAGE \cite{MuPara,mupage} and GENHEN \cite{simtools, simtools2} packages, respectively. For the simulation of atmospheric muons, the total amount of simulated events corresponds to 1/3 of the total livetime of the data set.
The Bartol flux \cite{bartol} is considered to represent the atmospheric neutrino flux. 
The simulation of the amount of optical background (\ce{^{40}K} and bioluminescence) is performed according to the collected data in order to account for the variations of the environmental conditions \cite{RBR}.

\begin{figure}[!ht]
\centering
        \includegraphics[width=.49\linewidth]{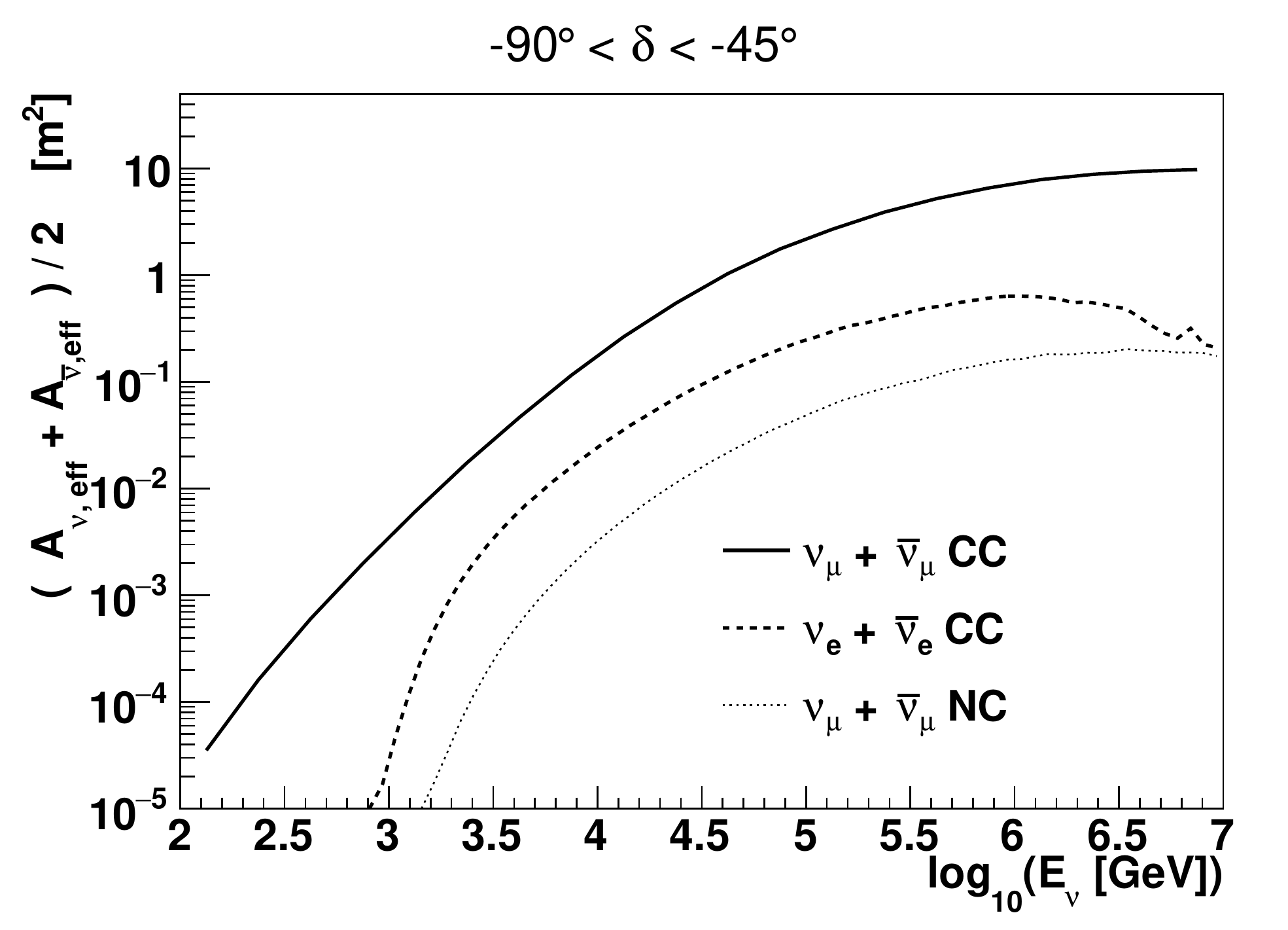}
        \includegraphics[width=.49\linewidth]{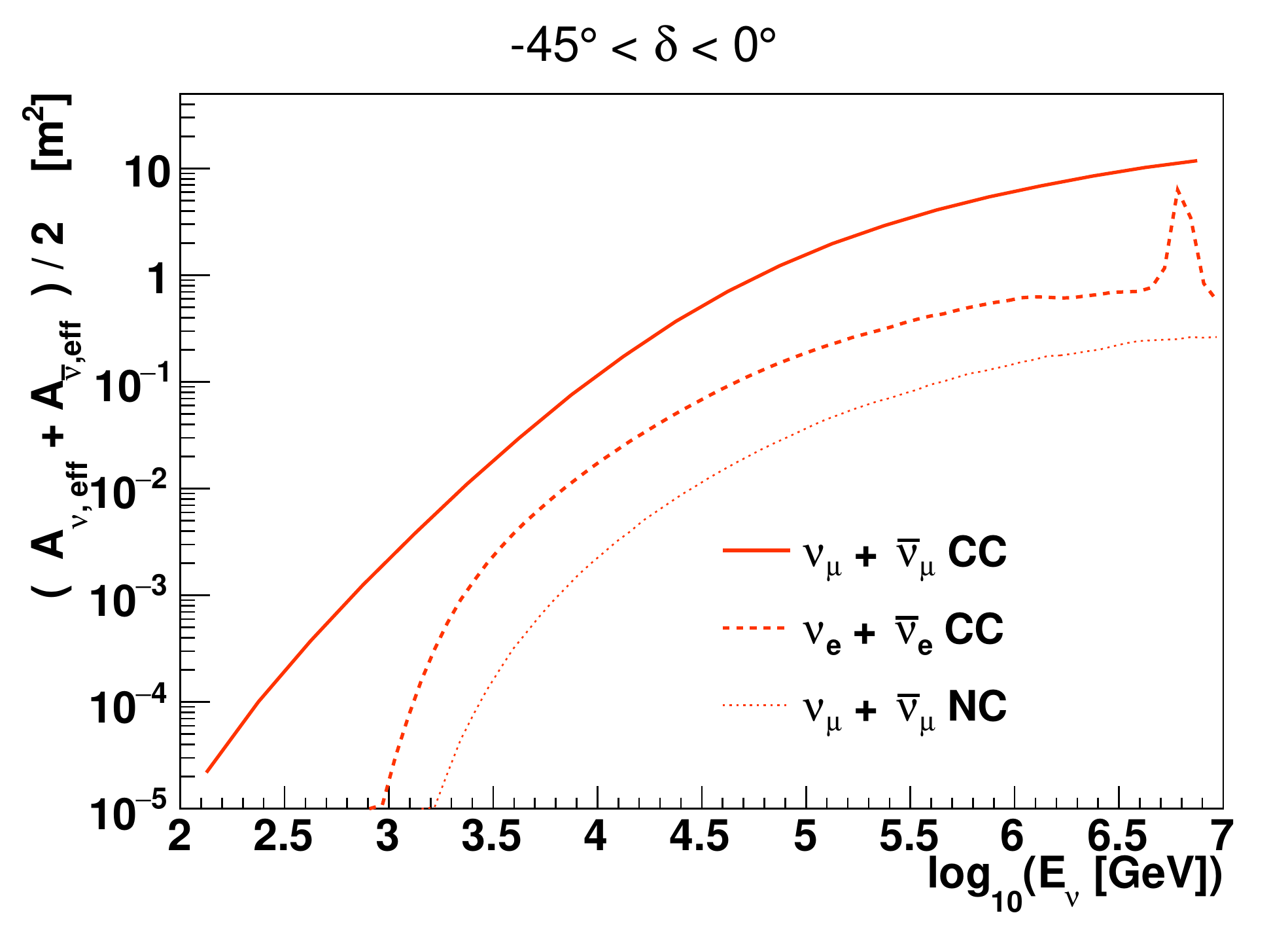}
        \includegraphics[width=.49\linewidth]{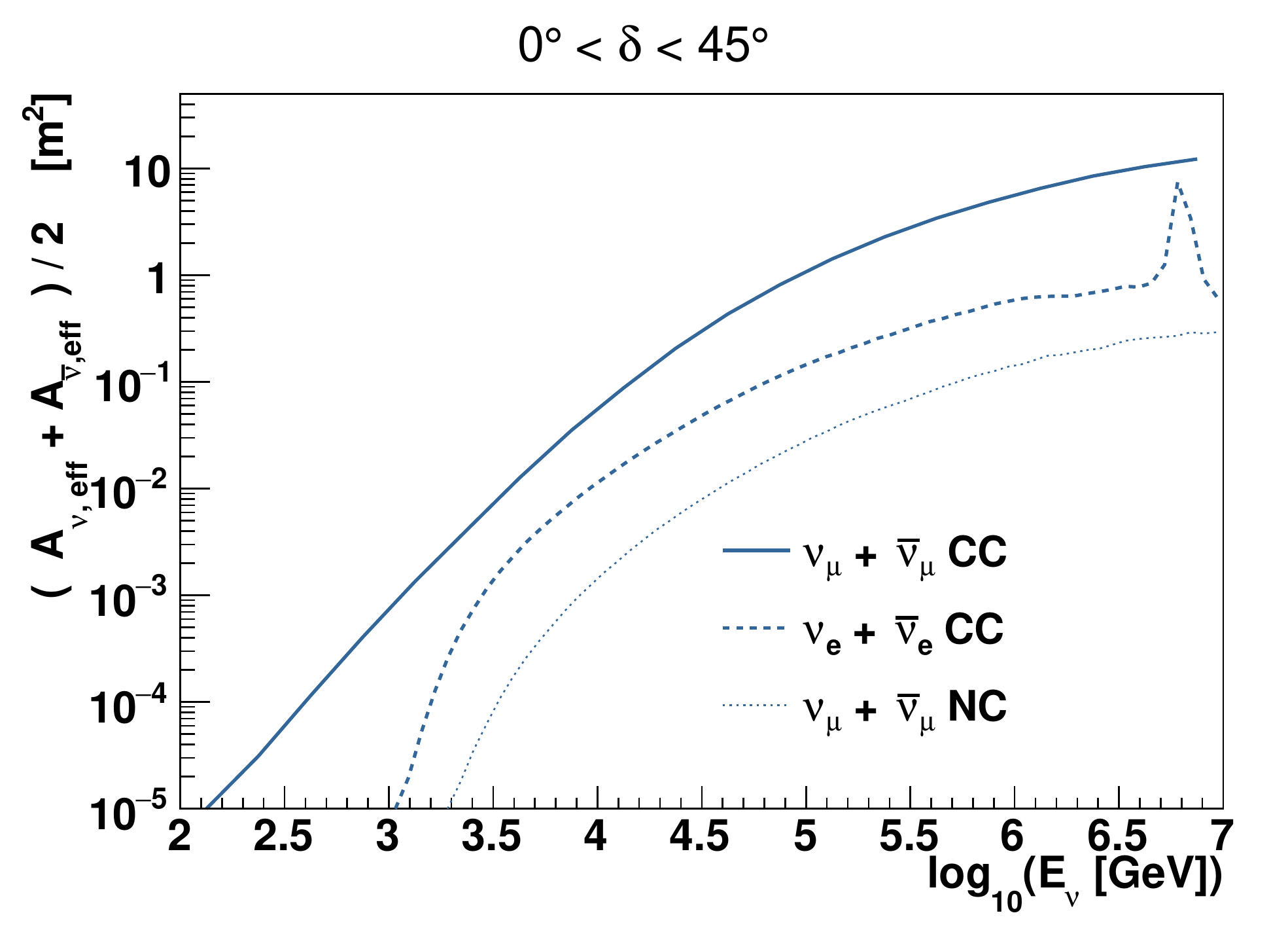}

	\caption{Effective area for ${\nu_\mu}$+$\bar{\nu}_\mu$ CC events after the track selection cuts (solid line) and for $\nu_{e}$+$\bar{\nu}_{e}$ CC and $\nu_\mu$+$\bar{\nu}_\mu$ NC events after the shower selection cuts (dashed lines) considering three declination ranges.}
    \label{fig:Aeff}
\end{figure}

The data used for the analysis was recorded between 2007 January 29th and 2015 December 31st. During this period, which includes the commissioning phase, the detector operation  with at least five lines corresponds to a total livetime of 2423.6 days. 
The event selection is optimized following a blind procedure on pseudo-data sets of data randomized in time (pseudo-experiments) before performing the analysis. A more detailed description of the pseudo-experiments can be seen in section \ref{sec:TS}.
The selection criteria for tracks and showers are explained in Sec. \ref{muon_sel} and \ref{show_sel}, respectively. These criteria have been optimized to minimize the neutrino flux needed for a $5\sigma$ discovery of a point-like source in $50\%$ of the pseudo-experiments. The effective area for $\nuan_{\mu}$\footnote{The notation $\nuan$ refers to both neutrinos and anti-neutrinos.} CC after the track selection cuts, and for $\nuan_{e}$ CC and $\nuan_{\mu}$ NC events after the shower selection cuts can be seen in Fig. \ref{fig:Aeff}.

The simulations produced to account for the interaction of $\nuan_{\tau}$ cover a limited livetime of the data. Because of this, the contribution due to the leptonic and hadronic channels after the interaction of $\nuan_{\tau}$ are estimated by scaling the contribution of other flavor neutrino events. A neutrino flux with an E$^{-2}$ energy is assumed in this scaling. The obtained rates are the following: to take into account the decay of the outgoing $\tau$ into a $\mu$ (branching ratio of $\sim$ 17\%) after a $\nuan_{\tau}$ Charged Current (CC) interaction, the number of events predicted by CC $\nuan_{\mu}$ interactions is increased by 9\%;
to take into account the decay of the outgoing $\tau$ into an electron (branching ratio of $\sim$ 17\%), the number of CC $\nuan_{e}$ is increased by 12\%; 
finally, to take into account the $\tau$ decaying into hadrons (branching ratio of $\sim$ 64\%) after a $\nuan_{tau}$ CC interaction and the $\nuan_{\tau}$ NC interactions, the small number of events predicted by $\nuan_{\mu}$ NC interactions is increased by 374\%.

%0.09 of the Charged Current (CC) $\nuan_{\mu}$ to take into account the decay of the outgoing $\tau$ into a $\mu$, 0.12 of the CC $\nuan_{e}$ to take into account the decay of the $\tau$ into an electron, and 3.74 of the NC $\nuan_{\mu}$, which takes into account the $\nu_\tau$ NC interaction and the $\tau$ decaying into hadrons after a CC interaction.

%The $\nuan_{\tau}$  contributions in the leptonic and hadronic channels are estimated by scaling the contribution of other flavor neutrino events assuming a neutrino flux with an $E^{-2}$ energy spectrum. The scaling has been calculated after performing simulations of $\nu_\tau$ events for a reduced livetime. The obtained rates are the following: 0.09 of the Charged Current (CC) $\nuan_{\mu}$ to take into account the decay of the outgoing $\tau$ into a $\mu$, 0.12 of the CC $\nuan_{e}$ to take into account the decay of the $\tau$ into an electron, and 3.74 of the NC $\nuan_{\mu}$, which takes into account the $\nu_\tau$ NC interaction and the $\tau$ decaying into hadrons after a CC interaction. 

\subsection{Muon track selection}
\label{muon_sel}
Muon tracks are reconstructed using a multi-step procedure that concludes with a maximum likelihood method \cite{antPS4y}. This likelihood takes into account the so-called \textit{hits}. A hit is defined as the digital information on the time and amplitude of a PMT signal, where the latter is proportional to the number of detected photons.
As in the previous publication \cite{lastPS}, muon events are selected applying cuts on the reconstructed zenith angle ($\cos\theta_{tr} > -0.1$), the estimated angular error ($\beta_{tr} < \unit{1}{\degree}$) and the parameter that describes the quality of the reconstruction ($\Lambda > -5.2$).
An approximated evaluation of the energy deposited per unit of path length is used to estimate the muon energy. An energy estimator, $\rho$, is defined using the hit charge, recorded by all PMTs used to reconstruct the track, and the length of the muon path in the detector \cite{dEdXcuts,dEdX}.
The energy estimator fails for events for which  the muon energy is below the value of the critical energy to produce significant energy losses due to radiative processes ($\sim$500 GeV), and for tracks with estimated track length, $L_\mu$, below \unit{380}{\metre}, yielding small values of $\rho$.
Such events are excluded from the analysis. Table~\ref{tab:TrackSel} gives an overview of the selection cuts applied for the simulated track sample. A total of 7622 neutrino candidates in the track channel are selected in data for this search. 

\begin{table*}[!h]
    \centering
    \caption{Selection cuts for the track sample and number of remaining simulated events after each step for atmospheric muons ($n_{\mu}^{atm}$), atmospheric neutrinos ($n_{\nu}^{atm}$) and cosmic neutrinos ($n_{\nu}^{E^{-2}}$) reconstructed as a track in the detector. For cosmic neutrinos, a flux according to $d\varPhi/dE = 10^{-8}\,(E / \si{GeV})^{-2} \, \si{GeV^{-1}\,cm^{-2}\,s^{-1}}$ is assumed. 
    \medskip }
    
    \label{tab:TrackSel}
    \begin{tabular}{l c c c c}
            Criterion      & Condition                                     & $n_{\mu}^{atm}$  & $n_{\nu}^{atm}$  & $n_{\nu}^{E^{-2}}$  \\
            \hline                                                              
            Trigger     &                                       & $4.9\times 10^{8}$ & $ 6.3 \times 10^{4}$ & 204  \\
            Up-going       & $\cos\theta_{tr} > -0.1$          & $4.3\times 10^{7}$ & $5.0\times 10^{4}$ & 151      \\
            Angular error estimate
                 &     $\beta_{tr} < \unit{1}{\degree}$         & $2.2\times 10^{7} $ & $3.3\times 10^{4}$ & 105     \\
            Track reconstruction quality & $\Lambda > -5.2$      & $1513$ & 7475 & 44    \\\smallskip                                                
            Track length and energy cut   & $ L_{\mu} > 380$ m, $\log_{10}(\rho) > 1.6$           & 1117 & 7086 & 41     \\
            
        \end{tabular}
%     }
\end{table*}

\subsection{Shower selection}
\label{show_sel}

Shower events are reconstructed with a new algorithm based on a two-step procedure. In the first step, the interaction vertex is obtained by the maximization of an M-estimator, $M_\mathrm{est}$, which depends on the time and charge of the hits. The direction of the event is estimated with a maximum likelihood method, using the information of the reconstructed interaction vertex and the detected amplitude of the OMs.
Shower events are required to be reconstructed as up-going or coming from close to the horizon ($\cos\theta_{sh} > -0.1$) with a restriction on the angular error estimate ($\beta_{sh} < \unit{30}{\degree}$). The interaction vertex of each event is also required to be reconstructed inside or close to the instrumented volume.
To further reduce the background from mis-reconstructed atmospheric muons, additional selection cuts are imposed. 
These cuts are based on the $M_\mathrm{est}$ value, on a Random Decision Forest classifier value, $RDF$, made with parameters provided by an alternative shower reconstruction \cite{Dusj-Paper}, and on a likelihood, $\mathscr L_\mathrm{\mu}$ or muon likelihood, that uses information of the hits in the event.
A description of these cuts is given in Appendix \ref{appendix}. Events passing the muon track selection are excluded from the shower channel making the two samples mutually exclusive. The full list of selection cuts is summarized in Table~\ref{tab:ShowSel}. The selection yields 180 shower events.  

\begin{table*}[!h]
    \centering 
    \caption{Selection cuts for the shower sample and number of remaining simulated events after each step for atmospheric muons ($n_{\mu}^{atm}$), atmospheric neutrinos ($n_{\nu}^{atm}$) and cosmic neutrinos ($n_{\nu}^{E^{-2}}$) reconstructed as a shower in the detector. For cosmic neutrinos, a flux according to $d\varPhi/dE = 10^{-8}\,(E / \si{GeV})^{-2} \,\si{GeV^{-1}\,cm^{-2}\,s^{-1}}$ is assumed. 
    Refer to Appendix \ref{appendix} for more details.\medskip }
    \label{tab:ShowSel}
%     \resizebox{\linewidth}{!}{ 
        \begin{tabular}{l c c c c c}
            Criterion      & Condition                                     & $n_{\mu}^{atm}$  & $n_{\nu}^{atm}$  & $n_{\nu}^{E^{-2}}$ \\	
            \hline                                                          
            Track Veto     & not selected as muon track                & $4.9 \times 10^{8}$ & $5.6 \times 10^{4}$ & 160   \\
            Up-going       & $\cos\theta_{sh} > -0.1$          & $1.5 \times 10^{8}$ & $2.3 \times 10^{4} $ & 90       \\
            Interaction vertex    & $R_{sh} < \unit{300}{\metre}$,                                                                                                   
                            $|Z_{sh}| < \unit{250}{\metre}$         & $7.7 \times 10^{7}$ & $2.1 \times 10^{4} $ & 80     \\
            M-estimator    & $M_\mathrm{est} < 1000$                       & $7.2 \times 10^{7}$ & $2.0 \times 10^{4} $ & 80      \\
            
            RDF       & $ RDF > 0.3$                           &  $8.0 \times 10^{4}$ & 2044 & 24     \\
            Muon likelihood      & $\mathscr L_\mathrm{\mu} > 50$           & 90 & 109 & 12     \\
        \end{tabular}
%     }
\end{table*}

\subsection{Comparison between data and simulations}

\begin{figure*}

        \includegraphics[width=0.48\linewidth]{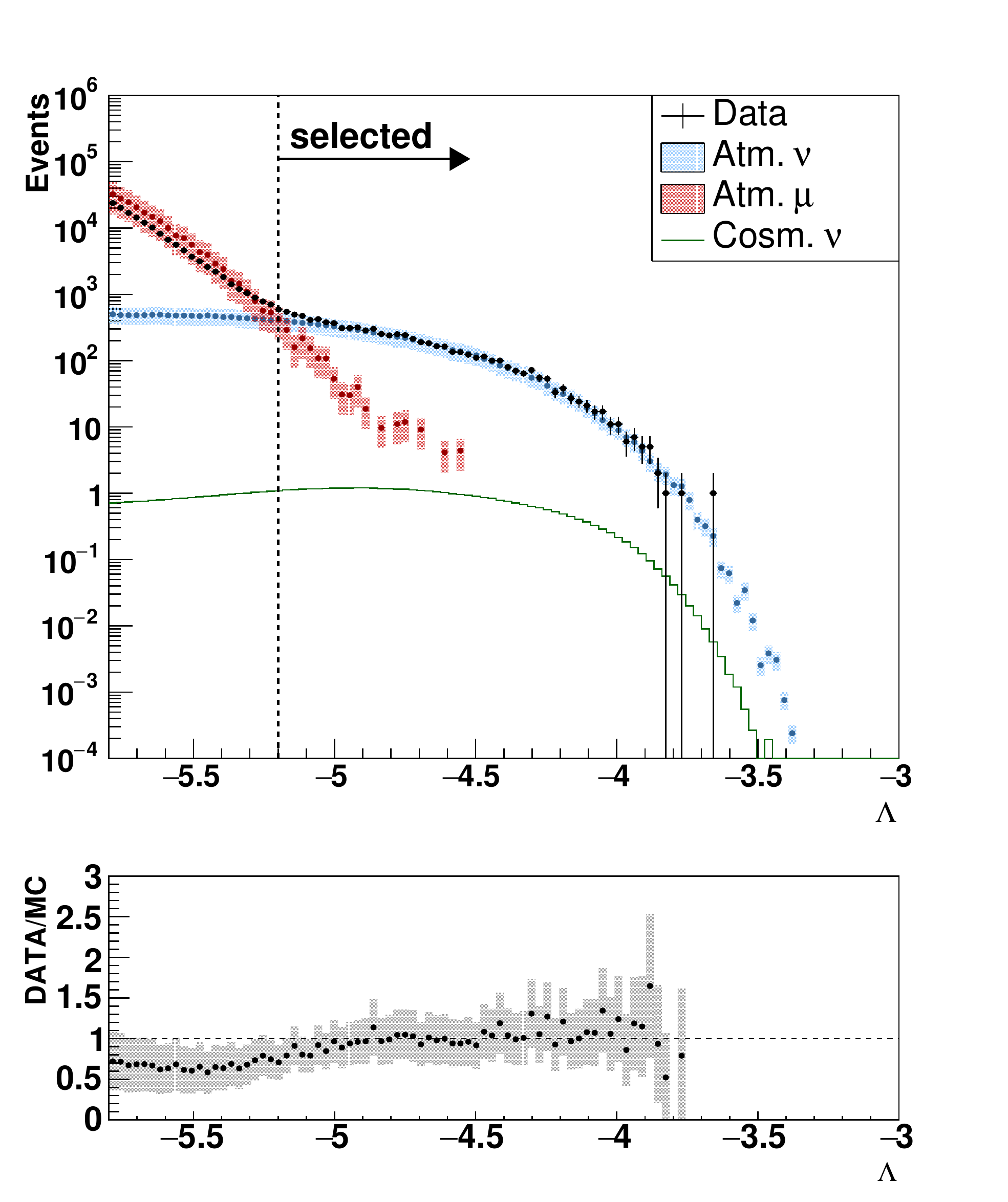}
        \includegraphics[width=0.48\linewidth]{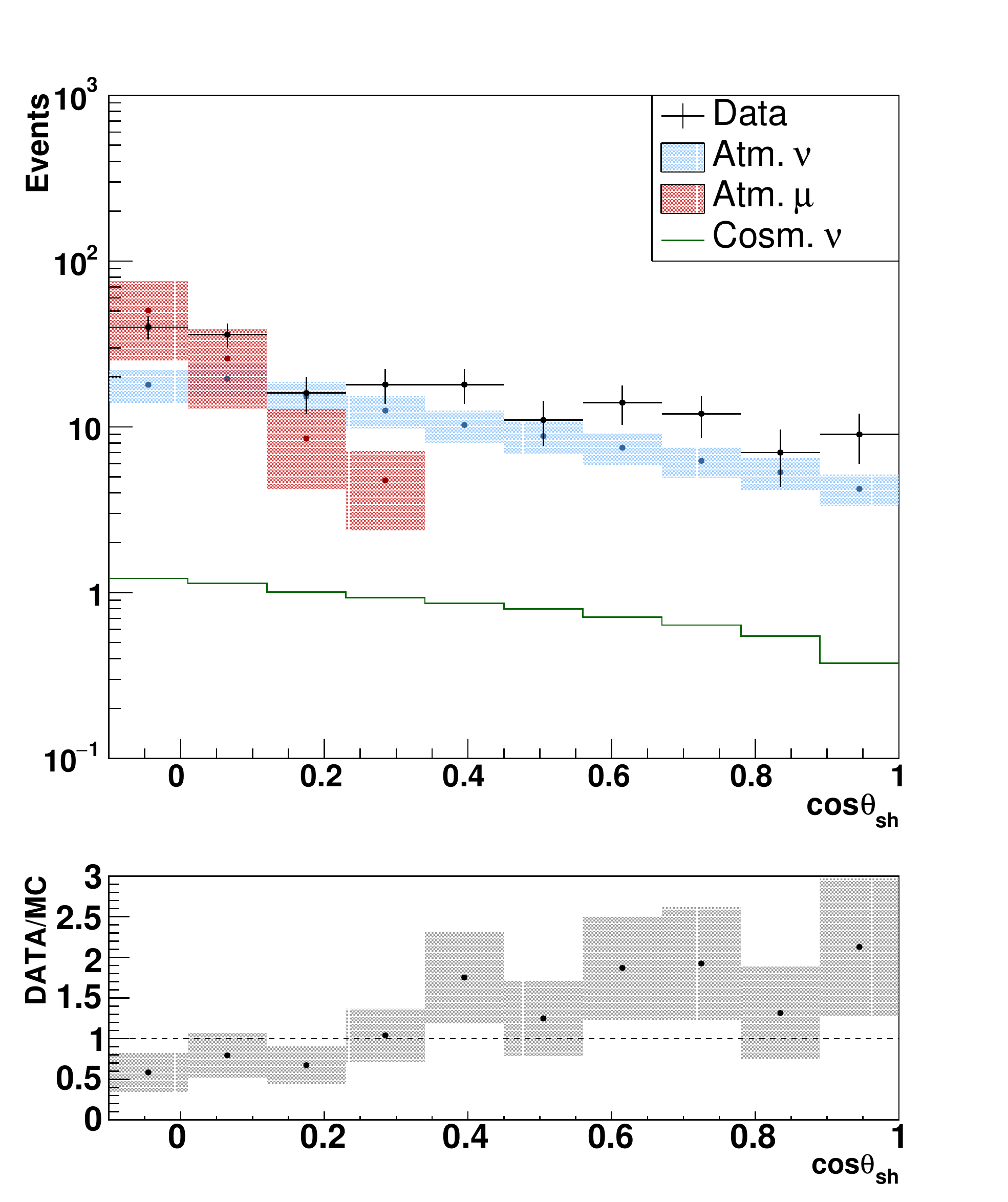}

	\caption{Comparison of the data with Monte Carlo (MC) simulations as a function of the quality parameter $\Lambda$ (left). This figure corresponds to the event distribution after a cut on the estimated angular error ($\beta_{tr} < \unit{1}{\degree}$) and on the reconstructed zenith angle ($\cos\theta_{tr} > -0.1$). The dashed vertical line marks the cut value. Right: comparison of the data with the simulations in the zenith $\theta_{sh}$ of the reconstructed shower direction. This figure corresponds to the event distribution after all shower selection cuts presented in Table~\ref{tab:ShowSel}. For the cosmic neutrinos, a flux according to $d\varPhi/dE = 10^{-8}\,(E / \si{GeV})^{-2} \, \si{GeV^{-1}\,cm^{-2}\,s^{-1}}$ is assumed in both figures. The two bottom plots show the data to MC ratio, where the number of MC events is the sum of atmospheric muons and neutrinos.}
     \label{fig:data_mc}
\end{figure*}

Figure~\ref{fig:data_mc}-left compares the distributions of the quality parameter $\Lambda$ for different types of simulated events with the data set  for the track channel. Figure~\ref{fig:data_mc}-right shows the comparison in the reconstructed zenith angle for the shower channel. It is estimated that about 13\% of the selected muon tracks and $52\,\%$ of the selected shower events are atmospheric muons mis-reconstructed as up-going.
 The simulation overestimates the number of events by $8\,\%$ ($17\,\%$) in the track (shower) channel for the final set of cuts. This difference is well within the overall systematic uncertainty on the atmospheric neutrino flux normalization \cite{dEdX}. A larger overestimation of events is observed in the region where the background of mis-reconstructed atmospheric muons is dominant.
 On the other hand, in the cascade channel, an underestimation of events is observed for zenith angles larger than $cos(\theta_{sh} ) >$ 0.4, which can be explained due to the large uncertainties of the atmospheric neutrino flux. The searches for an excess from a point-like direction in the sky are, compared to other searches, less influenced by higher levels of background contamination. Referring to the right panel of Fig. \ref{fig:data_mc}, even if there is a large contamination of atmospheric muons between -0.1 and 0.1, according to our pseudo-experiment simulations the current event selection produces better sensitivities compared to considering only events with $cos(\theta_{sh}) >$ 0.1.  A global good agreement between data and Monte Carlo justifies the procedure to optimize the  selection criteria for the separation of signal and background using simulated events.
 
 % On the other hand, in the cascade channel, an underestimation of events is observed for zenith angles larger than $cos(\theta_{sh}) >$ 0.4, which can be explained due to the large uncertainties of the atmospheric neutrino flux. Even if the amount of atmospheric muons is the main contribution for zenith angles smaller than $cos(\theta_{sh}) <$ 0.1, the calculated discovery flux is optimal for the current event selection. A global good agreement between data and Monte Carlo justifies the procedure to optimize the  selection criteria for the separation of signal and background using simulated events.%

\section{Search method}\label{sec:METHOD}
While atmospheric neutrino events are randomly distributed, neutrinos from point-like sources are expected to accumulate in spatial clusters. To find these clusters, a maximum likelihood ratio approach is followed. The likelihood used describes the data in terms of signal and background probability density functions (PDFs) and is defined as

\begin{align} \label{eq:pslik}
    \log \mathscr L_\mathrm{s+b} = &\sum_{\mathcal S} \sum_{i\in\mathcal S} \log \Big[ \mu_\mathrm{sig}^{\mathcal S}
     \mathscr {F}^{\mathcal S}_{i}  {\mathscr P}^{\mathcal S}_\mathrm{sig, i} + {\mathscr N}^{\,\mathcal S}  {\mathscr B}^{\,\mathcal S}_{i}  {\mathscr P}^{\mathcal S}_\mathrm{bkg, i}\Big]
    - \mu_\mathrm{sig} .
\end{align} 

In this equation, $\mathcal S$ denotes the sample ($\it{tr}$ for tracks, $\it{sh}$ for showers), $i$ indicates the event of the sample $\mathcal S$, $\mu_\mathrm{sig}^{\mathcal S}$ is the number of signal events fitted to in the $\mathcal S$ sample, $\mathscr {F}^{\mathcal S}_{i}$ is a parameterization of the point spread function, ${\mathscr {P}}^{\mathcal S}_\mathrm{sig, i}$ is derived from the probability density function of the energy estimator, yielding  the probability of measuring the signal with the reconstructed energy of the event $i$, ${\mathscr N}^{\,\mathcal S}$ is the total number of events in the $\mathcal S$ sample, ${\mathscr B}^{\,\mathcal S}_{i}$ is the background rate obtained from the distribution of the observed background events at the declination of event $i$, ${\mathscr {P}}^{\mathcal S}_\mathrm{bkg, i}$ is the probability density function of the energy estimator for background and $\mu_\mathrm{sig} = {\mu}^{tr}_\mathrm{sig} + {\mu}^{sh}_\mathrm{sig}$ is the total number of fitted signal events.
More details on the components of the PDFs are given below.

\subsection{Point spread function}
\begin{figure}[]
    \centering
    \includegraphics[width=.8\linewidth]{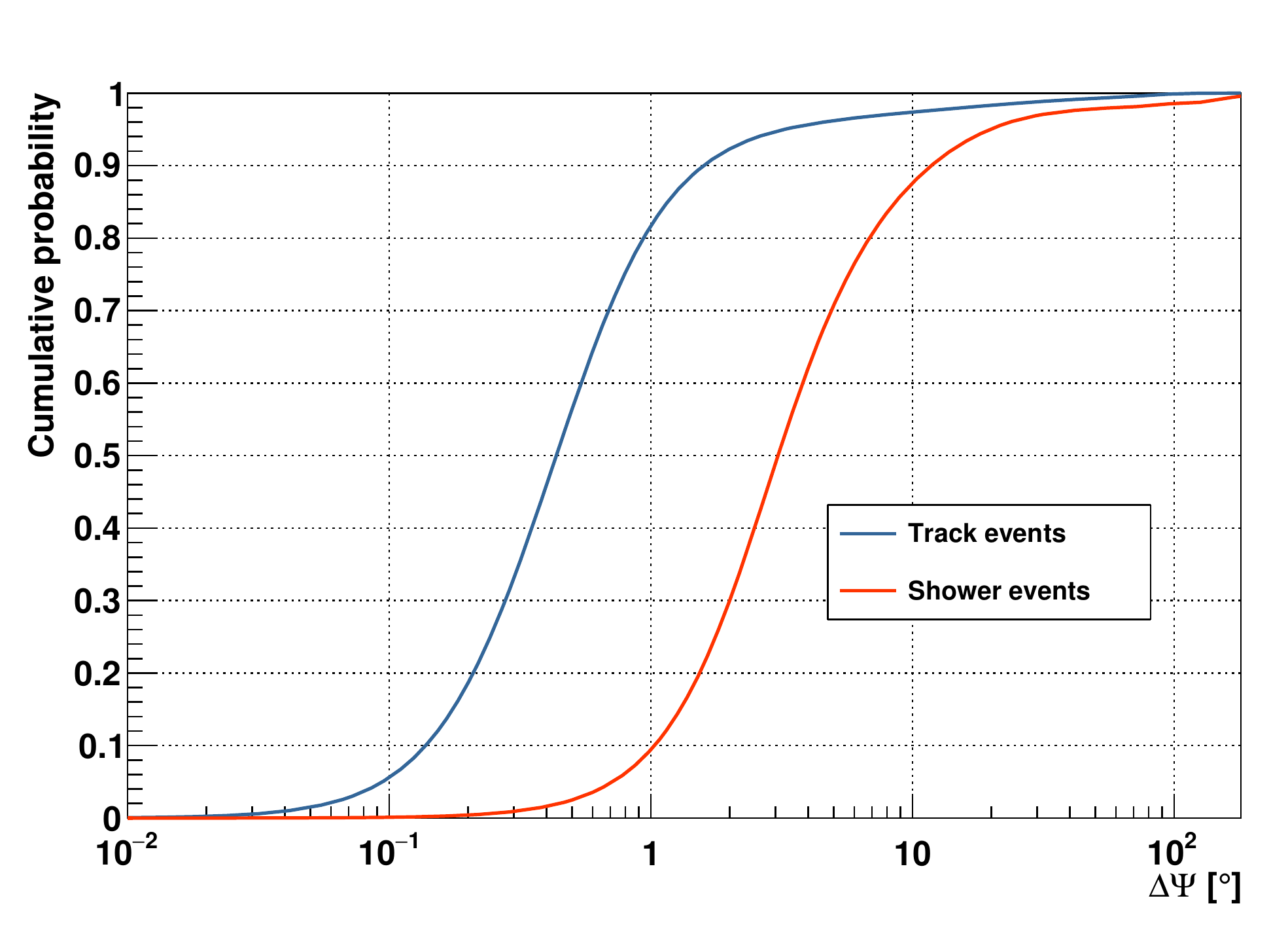}
    \caption{
    Probability to obtain a reconstructed angle within $\Delta\Psi$ between the reconstructed direction of track-like events (blue) and shower-like events (red) with respect to the true Monte Carlo neutrino direction. A neutrino flux with an $E^{-2}$ energy spectrum is assumed.
 	}
    \label{fig:psf_cum}
\end{figure}

The distribution of signal events around a hypothetical point-like source is described by the point spread function (PSF) $\mathscr F$. The PSF is defined as the probability density to find a reconstructed event at an angular distance $\Delta\Psi$ around the direction of the source. It depends on the angular resolution of the event sample. 
Figure~\ref{fig:psf_cum} shows the cumulative distributions of the angular distance between the reconstructed and true neutrino direction for track and shower events. The PSFs are determined from Monte Carlo simulations of neutrinos with an $E^{-2}$ energy spectrum.
The figure shows that about $50\,\%$ of the track (shower) events are reconstructed within \unit{0.4}{\degree} (\unit{3}{\degree}) of the parent neutrino. 

\subsection{Background rate}
\begin{figure*}[h]

        \includegraphics[width=0.48\linewidth]{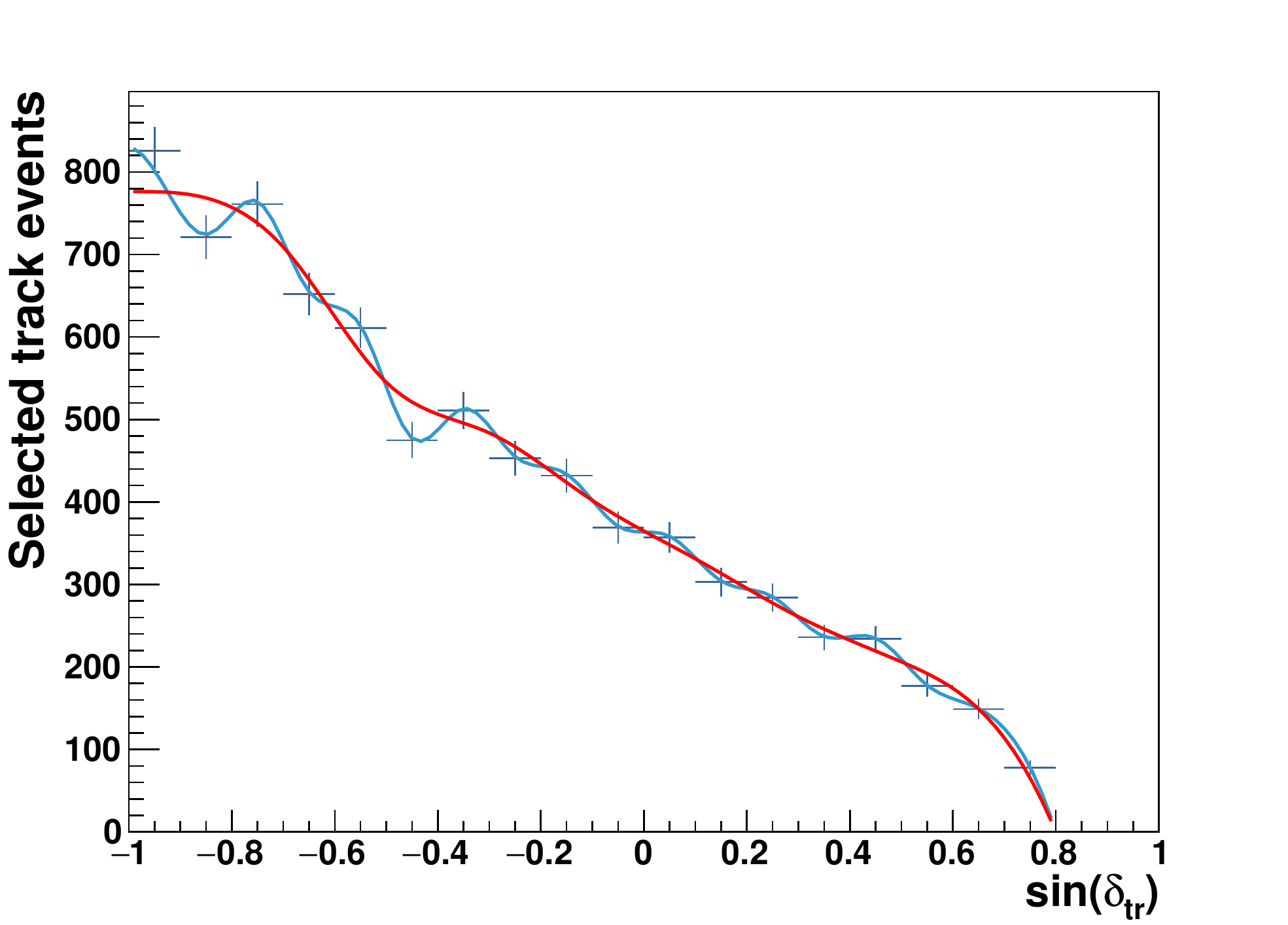}
        \includegraphics[width=0.48\linewidth]{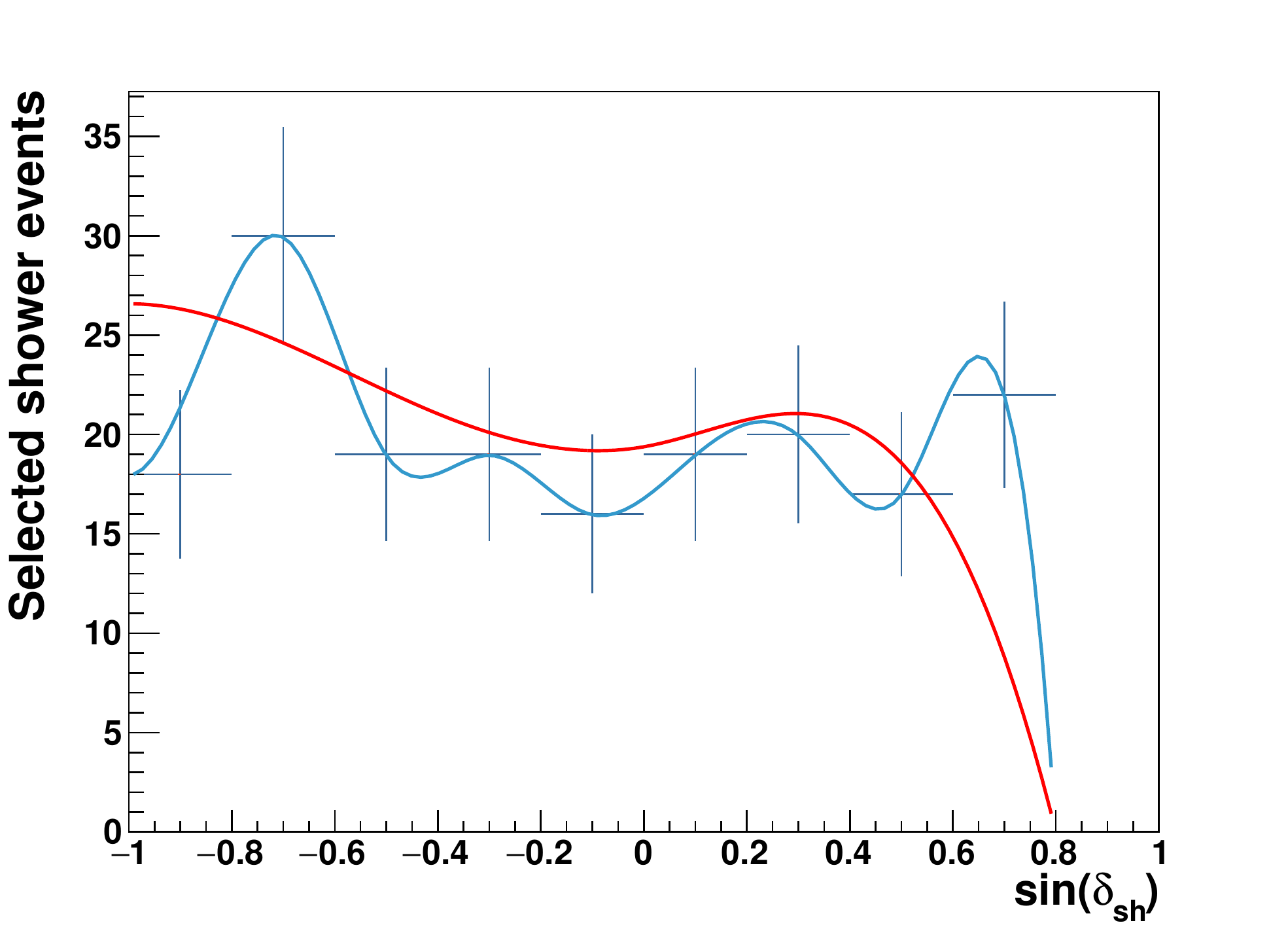}

    \caption{Number of selected track-like (left) and shower-like (right) events as a function of the reconstructed declination. The red and blue lines are different spline parametrisations (see Sec.~\ref{sec:SYST}). The different shape for showers is mainly due to a higher relative contamination of atmospheric muons in the sample.
    }
    \label{fig:background}
\end{figure*}

The background rate $\mathscr B$ is described as a function of the declination, $\delta$. Given the small expected contribution of a cosmic signal in the overall data set, the background rate is estimated directly from the measured data.
Due to the Earth's rotation and a sufficiently uniform exposure, the background is considered independent of right-ascension, $\alpha$.
The rate of selected events as a function of declination is shown in Fig.~\ref{fig:background}.

\subsection{Energy estimator}
Neutrinos generated in the atmosphere have a much softer energy spectrum  ($\propto E^{-3.7}$) than neutrinos from the expected astrophysical flux proportional to $E^{-2}$. 
For this reason, the energy estimator information is used in the likelihood to further distinguish between cosmic signal and atmospheric background. 

For the shower channel, the number of hits ($N_{sh}$) used by the reconstruction algorithm is employed as energy estimator. 

A different and more elaborate approach is assumed for the track channel. 
In this case the estimator $\rho$ is used as a proxy for the energy of the neutrino event. The information of the event angular error estimate $\beta_{tr}$ is also included. Moreover, the dependence of the energy estimator on the declination of the event is taken into account by generating both the signal and the background PDF in steps of 0.2 over $\sin\delta$. 

\subsection{Implementation}
\label{sec:TS}

The significance of any observation is determined by a test statistic denoted as $\mathcal Q$ which is defined from the likelihood as
\begin{equation}
    \mathcal Q = \log \mathscr L_\mathrm{s+b} - \log \mathscr L_\mathrm{b}.
    \label{eq:teststat}
\end{equation}
The $\mathcal Q$ distributions for different signal strengths are determined from pseudo-experiments. In these, $\BigO{10^{4}\sim10^{5}}$ random sky maps are generated with a number of background events that follow the declination-dependent event distribution as seen in the actual data, and a uniform right ascension. In addition, signal events are injected according to the investigated spectrum by assuming either a point or extended source profile. 
In equation~(\ref{eq:teststat}) $\mathscr L_\mathrm{b}$ corresponds to the definition of $\mathscr L_\mathrm{s+b}$ in equation~(\ref{eq:pslik}) evaluated with the same parameters as the maximum likelihood estimate but with the numbers of signal events set to zero: $\mu^\mathrm{tr}_\mathrm{sig} = \mu^\mathrm{sh}_\mathrm{sig} = 0$ (background-only case). 

\begin{figure}[h]
    \centering

    \includegraphics[width=0.7\linewidth]{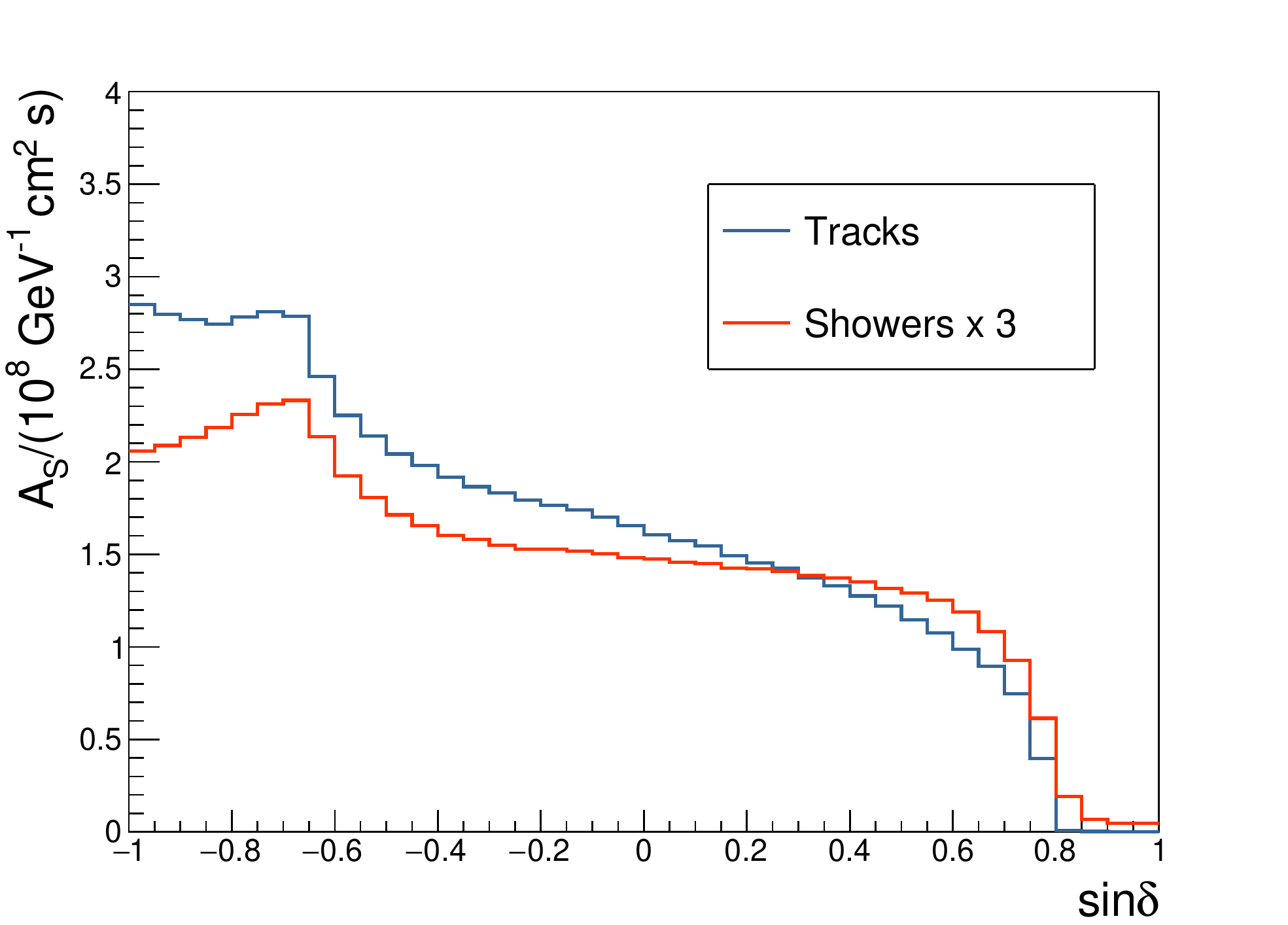}
    \caption{The acceptance as a function of the source declination for an $E^{-2}$ energy spectrum with a flux normalization factor of $\varPhi_0 = \unit{e-8}{\unitFluxNorm}$ for the track (blue) and shower (red) samples. For better visibility, the acceptance for showers is scaled up by a factor 3.}
    \label{fig:acceptance}
\end{figure}

In the likelihood maximization, the position in the sky of the fitted source is either kept fixed or allowed to be fitted within specific limits depending on the type of search (see Sec. \ref{sec:RESULTS}). Furthermore, the values of $\mu^\mathrm{tr}_\mathrm{sig}$ and $\mu^\mathrm{sh}_\mathrm{sig}$ are left free to vary, and can indeed go below zero to reflect the degree of absence of events around the probed coordinates.
The declination-dependent acceptance for a given sample, $A_{S}(\delta)$, is defined as the proportionality constant between a given flux normalization $\varPhi_0 = E^2 d\varPhi/dE$ and the expected number of signal events for this particular flux. It can be expressed in terms of the effective area, $\mathcal{A}_{\mathrm{eff},S}(E_\nu, \delta)$:
\begin{equation}
    A_{S}(\delta) = \varPhi_0^{-1} \iint dtdE_\nu\ \mathcal{A}_{\mathrm{eff}, S}(E_\nu, \delta) \frac {d\varPhi} {dE_\nu},
\end{equation}
where the integral is over the livetime of all selected runs (2423.6 days) and over an energy range large enough to include all potential events within the sensitivity of ANTARES. With the assumed $E^{-2}$ spectrum, 90\% of the events are found in an energy range between 2$\cdot 10^{3}$ and 3$\cdot 10^{6}$ GeV for the track channel (between 5$\cdot 10^{3}$ and 4$\cdot 10^{6}$ GeV  for the shower channel). 
Figure~\ref{fig:acceptance} shows how the acceptances for tracks and showers depend on the declination. 

\section{Search for neutrino sources}\label{sec:RESULTS}

The search for astrophysical neutrino sources presented in this paper is performed with four approaches.
\begin{description}  
    \item [1] {\it Full sky search.} In the first method, the whole visible sky of ANTARES is scanned in an unbinned way to search for spatial clustering of events with respect to the expected background. 	
    \item [2] {\it Candidate list search.} In the second approach, the directions of a pre-defined list of known objects which are neutrino source candidates are investigated to look for an excess or (in the case of null observation) to determine an upper limit on their neutrino fluxes.
    \item[3] {\it Galactic Center region.} The third search is similar to the full sky search but  restricted to a region centered in the origin of the galactic coordinate system ($\alpha$, $\delta$) = (266.40$^\circ$,--28.94$^\circ$) and defined by an ellipse with a semi-axis of 15$^\circ$ in the direction of the galactic latitude and a semi-axis of 20$^\circ$ in galactic longitude. 
The motivation relies on the number of high-energy neutrino events observed by the IceCube (IC) detector \cite{IC3years, IC4yproc} that appear to cluster in this region. Furthermore, the HESS Collaboration recently discovered an accelerator of PeV protons in the Galactic Center \cite{HESSPeV} that could produce high-energy neutrinos. 
   	\item[4] {\it Sagittarius A*.} Finally, the fourth approach tests the location of Sagittarius A* as an extended source by assuming a Gaussian emission profile of various widths.
	\end{description}

\begin{figure}[h]
    \centering
	\includegraphics[width=1.0\linewidth]{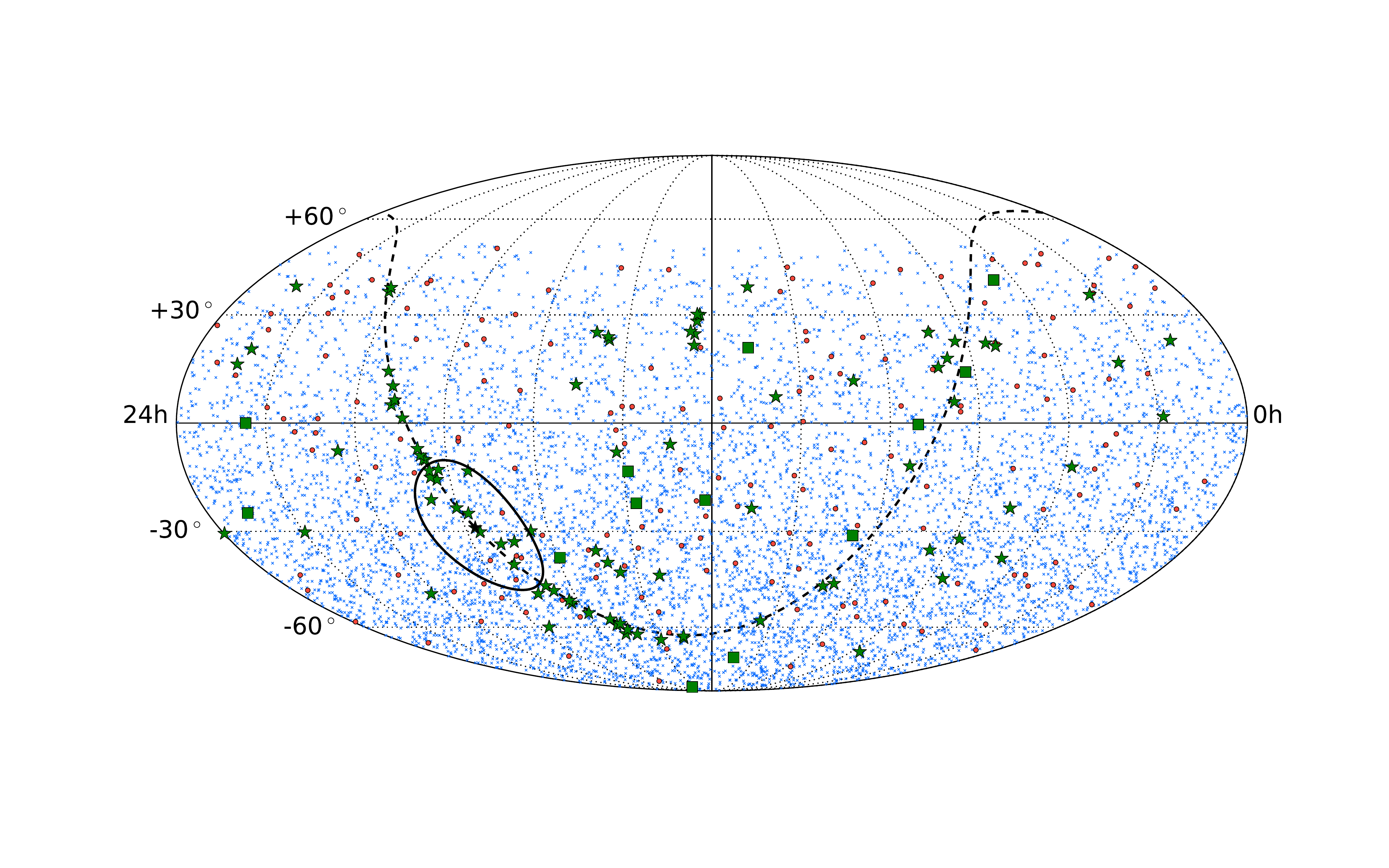}
    \caption{Sky map in equatorial coordinates of the 7622 track (blue  crosses) and the 180 shower (red circles) events passing the selection cuts. Green stars indicate the location of the 106 candidate neutrino sources, and green squares indicate the location of the 13 considered tracks from the IceCube high energy sample events or HESE (see Sec. \ref{sec:CL}). The black solid ellipse indicates the search region around the Galactic Center, in which the origin of the galactic coordinates is indicated with a black star. The black dashed line indicates the galactic equator.}
    \label{fig:SkyMap}
\end{figure}

Figure \ref{fig:SkyMap} represents the event sample in equatorial coordinates in the ANTARES visible sky. The considered neutrino source candidates and the search region around the Galactic Center are also indicated.

\subsection{Full sky search}
\label{fullsky}

\begin{figure}[h]
    \centering
   \includegraphics[width=\linewidth]{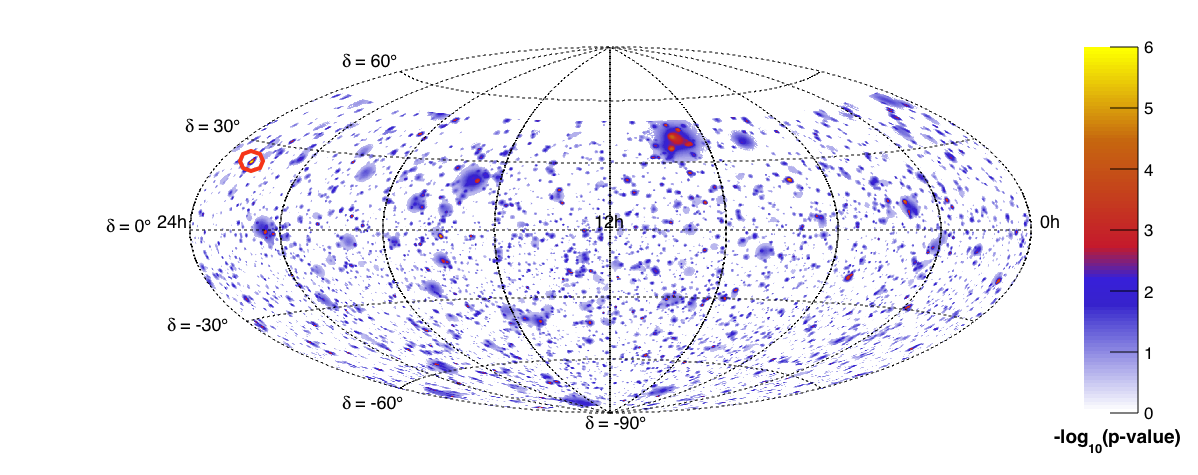}
   \caption{%
     Sky map in equatorial coordinates of pre-trial p-values for a point-like source of the ANTARES visible sky. The red circle indicates the location of the most significant cluster of the full sky search. For this map, a smaller grid size of 0.2$^\circ \times$ 0.2$^\circ$ was used.
     }
  \label{fig:Pretrial}
\end{figure}

\begin{figure}[p]
    \centering
    \includegraphics[width=0.75\textwidth]{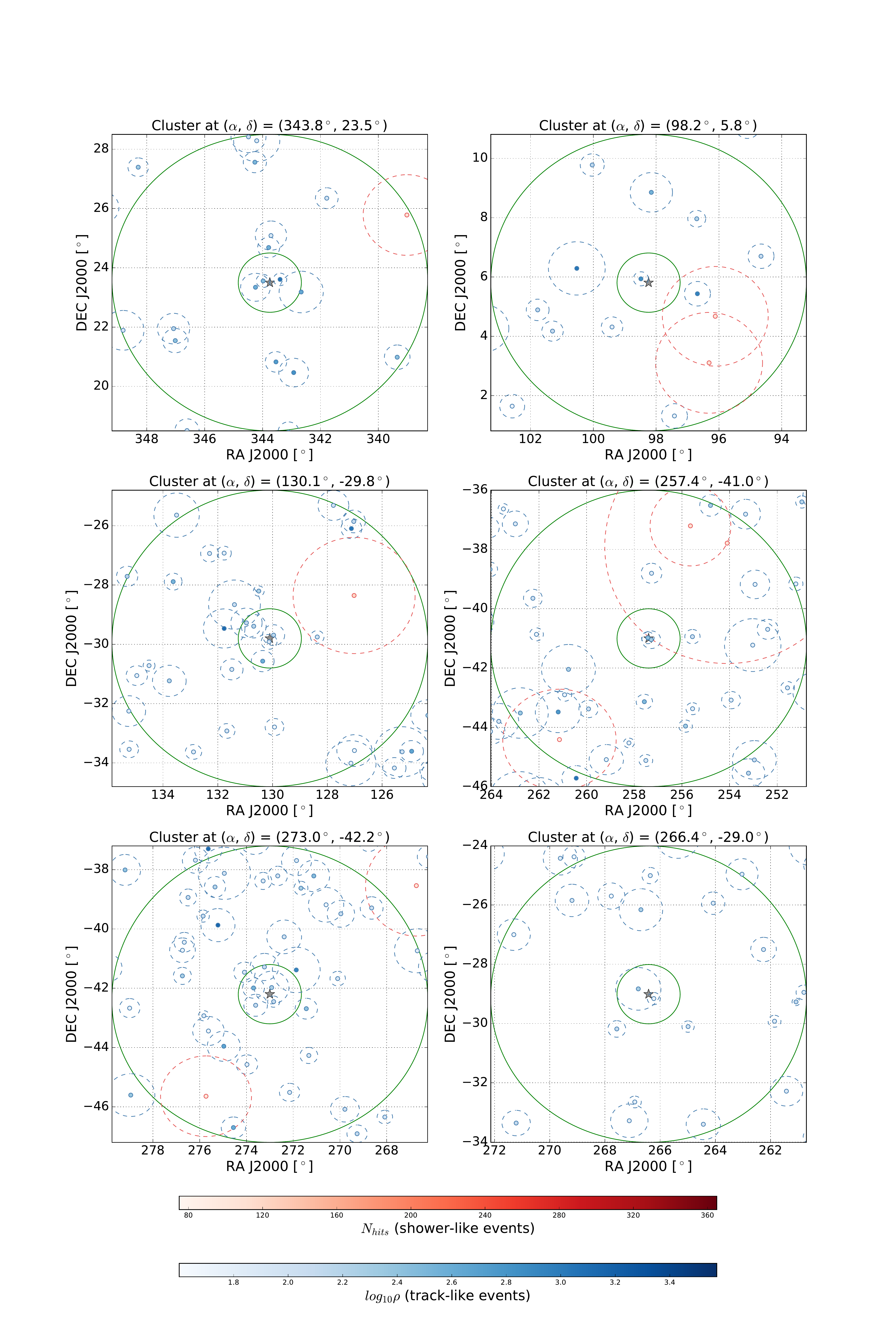}
    \caption{\scriptsize Distribution of events in the ($\alpha$,  $\delta$) (RA, DEC) coordinates for the most significant clusters found in the full sky search (top left), candidate list search (HESSJ0632+057) (top right), search over the track events from the IceCube HESE sample (track with ID = 3) (middle left), search around the Galactic Center for an $E^{-2}$ point-like source (middle right), search around the Galactic Center for an $E^{-2.5}$ point-like source (bottom left) and at the location of Sagittarius A* (bottom right). In all figures, the inner (outer) green line depicts the one (five) degree distance from the position of the best fit or known location, indicated as a gray star. The red points denote shower-like events, whereas the blue points indicate track-like events. Different tones of red and blue correspond to the values assumed by the energy estimators: the number of hits (shower-like events) and the $\rho$ parameter (track-like events) as shown in the legend. The dashed circles around the events indicate the angular error estimate. }
  \label{fig:AllClusters}
\end{figure}

\begin{figure}[h!]
    \centering
    \includegraphics[width=1\linewidth]{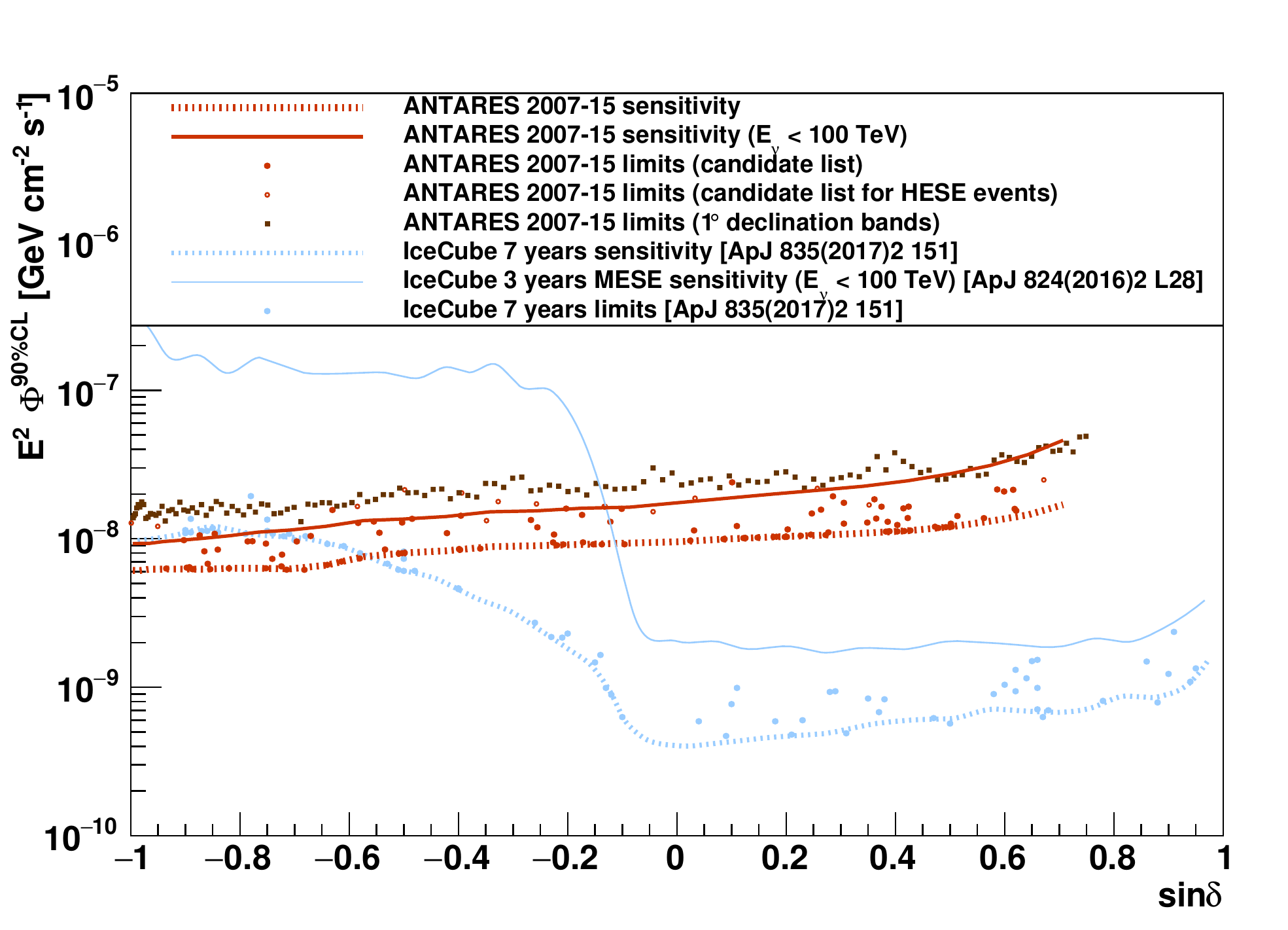}
    \caption{Upper limits at a $90\,\%$ C.L. on the signal flux from the investigated candidates assuming an $E^{-2}$ spectrum (red circles). The dashed red line shows the ANTARES sensitivity (defined as the median upper limit at 90\% C.L. for a background-only case) and the blue dashed line the sensitivity of the seven years point-like source analysis by the IceCube Collaboration for comparison \cite{icecubePS}. The upper-limits obtained in this analysis are also included (blue dots). The ANTARES 5$\sigma$ pre-trial discovery flux is a factor 2.5 to 2.9 larger than the sensitivity.  The curve for the sensitivity for neutrino energies under 100 TeV is also included (solid red line). The IceCube curve for energies under 100 TeV (solid blue line) is obtained from the 3 years MESE analysis \cite{icecubeMESE}. The limits of the most significant cluster obtained in bands of 1$^\circ$ in declination (dark red squares) are also shown.}
    \label{fig:LimitsFix}
\end{figure}

In the full sky search, the whole visible sky of ANTARES is divided on a grid with boxes of $\unit{1}{\degree} \times \unit{1}{\degree}$ in right ascension and declination for the evaluation of the $\mathcal Q$-value defined in Eq.~(\ref{eq:teststat}). This value is maximised in each box by letting the location of the fitted cluster free between the $\unit{1}{\degree} \times \unit{1}{\degree}$  boundaries. Since an unbinned search is performed, events outside the grid boxes are indeed considered in each $\mathcal Q$-value maximisation. 
The pre-trial p-value of each cluster is calculated by comparing the $\mathcal Q$-value obtained at the location of the fitted cluster with the background-only $\mathcal Q$ obtained from simulations at the corresponding declination.
Figure~\ref{fig:Pretrial} shows the position of the cluster and the pre-trial p-values for all the directions in the ANTARES visible sky.
The most significant cluster of this search is found at a declination of $\delta = \unit{23.5}{\degree}$ and a right-ascension of $\alpha = \unit{343.8}{\degree}$ and  with a pre-trial p-value of $\num{3.84e-6}$.
To account for trial factors, this pre-trial p-value is compared to the distribution of the smallest p-values found anywhere in the sky when performing the same analysis on many pseudo-data sets. It is found that 5.9\% of pseudo-experiments have a smaller p-value than the one found in the final sample, corresponding to a post-trial significance of $1.9\sigma$ (two-sided convention). The upper limit on the neutrino flux coming from this sky location is $E^2 d\varPhi/dE = \unit{3.8e-08}{GeV\,cm^{-2}\,s^{-1}}$. 
The location of this cluster is found at a distance of 1.1$^\circ$ from event ID 3 of the 6 year Northern Hemisphere Cosmic Neutrino flux sample of IceCube \cite{IC6years_HE}. A rough estimate of the significance of this coincidence is performed. 26 out of the 29 of these events are found in the declination range between -5$^\circ$ and 30$^\circ$. The remaining events were excluded since the event density in the selected region is larger, and therefore the estimation is slightly more conservative. By assuming a random distribution of 26 events within this declination range, the probability of a random coincidence within 1$^\circ$ between at least one event and the most significant cluster of the full sky search is $\sim$1\%.
%The location of this cluster is found at a distance of 1.1$^\circ$ from event ID 3 from the 6 year Northern Hemisphere Cosmic Neutrino flux sample from IceCube \cite{IC6years_HE}. A rough estimate to obtain if this coincidence is significant enough is performed. 26 out of the 29 of these events are found in a declination range between -5$^\circ$ and 30$^\circ$. By assuming a random distribution of 26 events within this declination range, a random coincidence within 1$^\circ$ between at least one event and the most significant cluster of the full sky search is $\sim$1\%. 
The distribution of events of this cluster is shown in Fig.~\ref{fig:AllClusters}-top-left. It contains 16(3) tracks within $\unit{5}{\degree}(\unit{1}{\degree})$ and 1 shower event within $\unit{5}{\degree}$. The upper limits of the highest significant cluster in bands of 1$^\circ$ in declination at a 90\% Confidence Level (C.L.) obtained using the Neyman method \cite{neyman} are shown in Fig. \ref{fig:LimitsFix}. The limits computed in this analysis are set on the one-flavor neutrino flux assuming equipartition at Earth of the three neutrino flavors.
%The location of this cluster is found at a distance of 1.1$^\circ$ from event ID 3 from the 6 year Northern Hemisphere Cosmic Neutrino flux sample from IceCube \cite{IC6years_HE}. 26 out of the 29 of these events are found in a declination range between -5$^\circ$ and 30$^\circ$. By assuming a random distribution of 26 events within this declination range, a random coincidence within 1$^\circ$ between at least one event and the most significant cluster of the full sky search is $\sim$1\%. The distribution of events of this cluster is shown in Fig.~\ref{fig:AllClusters}-top-left. It contains 16(3) tracks within $\unit{5}{\degree}(\unit{1}{\degree})$ and 1 shower event within $\unit{5}{\degree}$. The upper limits of the highest significant cluster in bands of 1$^\circ$ in declination at a 90\% Confidence Level (C.L.) obtained using the Neyman method \cite{neyman} are shown in Fig. \ref{fig:LimitsFix}.

\subsection{Candidate list}
\label{sec:CL}

The candidate list used in the last ANTARES point-like source analysis \cite{lastPS} contained neutrino source candidates both from Galactic and extra-Galactic origin listed in the TeVCat catalog \cite{tevcat}. These sources had been observed by gamma-ray experiments before July 2011 in the 0.1--100 TeV energy range and with declinations lower than $\unit{20}{\degree}$. Furthermore, since the energy of high energy gamma-rays of extra-galactic origin can degrade before they reach the Earth, extra-Galactic candidates were selected also among the sources observed by gamma-ray satellites in the 1--100 GeV energy range. 
This paper updates the neutrino search for the 50 objects considered in \cite{lastPS} with additional 56 galactic and extragalactic sources. The newly considered sources include those detected in the 0.1--100 TeV energy range by gamma-ray experiments after July 2011 and some bright sources with declinations between 20$^\circ$ and 40$^\circ$ not considered in the past. Additionally, the reconstructed direction of the IceCube multi-PeV track event \cite{IC6years_HE} and the 2HWC sources which are not coincident with any known source \cite{HAWC} have been included.
Finally, seven more sources are added: the three blazars with highest intensity observed by the TANAMI Collaboration that coincide with three events from the IceCube HESE sample \cite{tanami, tanami2, antares_blazars}, and the four gravitationally lensed Flat Spectrum Radio Quasars with the highest magnification factor analyzed in a previous work \cite{quasar}.

The list of the astronomical candidates is shown in Table~\ref{tab:LimitsFix} along with their equatorial coordinates, fitted number of signal events and upper limits on the flux. 

The most signal-like cluster is found at the location of HESSJ0632+057 at $(\alpha,\delta) = (\unit{98.24}{\degree},\unit{5.81}{\degree})$, with a pre-trial p-value of 0.16\%. The second and third most significant sources correspond to PKS1440-389 and PKS0235+164, with pre-trial p-values of 0.5\% and 5\%, respectively. To account for trial factors, the search is performed on the same list of sources using pseudo data-sets, from which the distribution of the smallest p-value for a background-only case is obtained. It is found that 13\% of the pseudo-experiments have a smaller p-value for any source compared to the one obtained for this location, corresponding to a post-trial significance of $1.5\sigma$ (two-sided convention). The cluster contains 11(1) tracks within $\unit{5}{\degree}(\unit{1}{\degree})$ and 2 shower events within $\unit{5}{\degree}$ around the source candidate.
The distribution of events around this source is shown in Fig.~\ref{fig:AllClusters}-top-right.
The sensitivities and limits calculated with the Neyman method at a 90\% C.L. and the 5$\sigma$ discovery flux for this search (assuming an $E^{-2}$ spectrum) are shown in Fig.~\ref{fig:LimitsFix} as a function of the declination. To prevent undesired effects of the Neyman construction \cite{Neyman-problems}, the maximum between the sensitivity (i.e., the median upper limit at a 90\% C.L. for a background-only case) and the limit for the particular location of the source are reported, as is customary in the field \cite{lastPS, icecubePS}.

The 13 track candidates from the IceCube HESE sample classified as muon tracks \cite{IC3years,  IC4yproc} are considered in a separate candidate list search. 
Since those events have a non-negligible angular error estimate, the direction parameters are not fixed but fitted within a cone of twice their angular error estimate around the direction given by the IceCube tracks.
The coordinates of these events are shown in Table~\ref{tab:LimitsHESE} together with their angular uncertainty (provided by the IceCube Collaboration), fitted number of signal events and upper limits on the flux derived from this analysis.

The muon track candidate from the HESE sample with the largest excess in fitted signal is the IceCube track with ID 3 and $\mu_\mathrm{sig} =5.3$. 
The fitted cluster is located at $(\alpha,\delta) = (\unit{130.1}{\degree},\unit{-29.8}{\degree})$, which is at a distance of 1.5$^\circ$ from the original HESE track at $(\alpha,\delta) = (\unit{127.9}{\degree},\unit{-31.2}{\degree})$. The observed post-trial p-value is 20\% (significance of 1.2$\sigma$). The upper limit on the signal from this candidate is $\varPhi_0^{90\,\%} = \unit{2.1e-8}{\unitFluxNorm}$.
The cluster is shown in Fig.~\ref{fig:AllClusters}-middle-left.

\setlength{\tabcolsep}{.3em}

\begin{table*}[p]
    \caption{\scriptsize List of astrophysical objects used in the candidate list search. Presented are the object's coordinates in declination ($\delta$) and right-ascension ($\alpha$). The first column reports the type of source: \textit{Binary} means X-Ray binary, \textit{GC} means Galactic Center, \textit{Radio} means Radio Galaxy, \textit{Sey2} means Seyfert 2 Galaxy,    \textit{UNID} means unidentified. The last two columns show the sum of the fitted number of signal track and shower events $\mu_\mathrm{sig} = {\mu}^{tr}_\mathrm{sig} + {\mu}^{sh}_\mathrm{sig}$, and the $90\,\%$ C.L. upper limits on the flux normalization factor $\varPhi_0^{90\,\%}$ (in units of $10^{-8}$ GeV cm$^{-2}$ s$^{-1}$). Candidates of the same type are sorted by declination. 
  \medskip}
    \label{tab:LimitsFix}
    \resizebox{\linewidth}{!}{
        \begin{minipage}{1.05\textwidth}
        \centering
 {\def\arraystretch{0.54}
    \footnotesize
    \begin{tabular}{l crrcc | l crrcc}
   
            Type & Name & $\delta [\si{\degree}]$   &    $\alpha [\si{\degree}]$    & $\mu_\mathrm{sig}$  & $\varPhi_0^{90\,\%}$ & Type & Name & $\delta [\si{\degree}]$   &    $\alpha [\si{\degree}]$    & $\mu_\mathrm{sig}$  & $\varPhi_0^{90\,\%}$     \\
\hline
BLLac & PKS2005-489 & -48.82 & 302.37 & 0.3 & 0.93 &  & PKS1406-076 & -7.90 & 212.20 & -- & 0.92 \\
 & PKS0537-441 & -44.08 & 84.71 & 0.6 & 0.96 &  & QSO2022-077 & -7.60 & 306.40 & 1.0 & 1.64 \\
 & PKS1440-389 & -39.14 & 220.99 & 2.9 & 1.56 &  & 3C279 & -5.79 & 194.05 & 0.8 & 1.59 \\
 & PKS0426-380 & -37.93 & 67.17 & -- & 0.70 &  & B1030+074 & 7.19 & 158.39 & -- & 1.01 \\
 & PKS1454-354 & -35.67 & 224.36 & 1.2 & 1.28 &  & PKS1502+106 & 10.52 & 226.10 & -- & 1.03 \\
 & TXS1714-336 & -33.70 & 259.40 & 0.8 & 1.31 &  & 3C454.3 & 16.15 & 343.50 & -- & 1.10 \\
 & PKS0548-322 & -32.27 & 87.67 & -- & 0.85 &  & 4C+21.35 & 21.38 & 186.23 & -- & 1.37 \\
 & H2356-309 & -30.63 & 359.78 & -- & 0.79 &  & B1422+231 & 22.93 & 216.16 & -- & 1.12 \\
 & PKS2155-304 & -30.22 & 329.72 & -- & 0.80 &  & PKS1441+25 & 25.03 & 220.99 & -- & 1.38 \\
 & 1ES1101-232 & -23.49 & 165.91 & -- & 0.85 & Radio & PKS0625-35 & -35.49 & 96.78 & -- & 0.74 \\
 & 1ES0347-121 & -11.99 & 57.35 & -- & 0.92 & SNR & LHA120-N-157B & -69.16 & 84.43 & -- & 0.63 \\
 & RGBJ0152+017 & 1.79 & 28.17 & -- & 1.14 &  & RCW86 & -62.48 & 220.68 & -- & 0.62 \\
 & RBS0723 & 11.56 & 131.80 & -- & 1.03 &  & MSH15-52 & -59.16 & 228.53 & -- & 0.68 \\
 & PKS0235+164 & 16.61 & 39.66 & 2.1 & 1.93 &  & SNRG327.1-01.1 & -55.08 & 238.65 & -- & 0.63 \\
 & RGBJ2243+203 & 20.35 & 340.98 & -- & 1.29 &  & RXJ0852.0-4622 & -46.37 & 133.00 & -- & 0.65 \\
 & VERJ0521+211 & 21.21 & 80.44 & 1.2 & 1.84 &  & RXJ1713.7-3946 & -39.75 & 258.25 & -- & 0.67 \\
 & S20109+22 & 22.74 & 18.02 & -- & 1.30 &  & W28 & -23.34 & 270.43 & 0.8 & 1.43 \\
 & PKS1424+240 & 23.79 & 216.75 & -- & 1.12 &  & SNRG015.4+00.1 & -15.47 & 274.52 & 0.2 & 1.34 \\
 & MS1221.8+2452 & 24.61 & 186.10 & -- & 1.13 &  & W44 & 1.38 & 284.04 & -- & 0.97 \\
 & 1ES0647+250 & 25.05 & 102.69 & -- & 1.65 &  & HESSJ1912+101 & 10.15 & 288.21 & -- & 1.03 \\
 & S31227+25 & 25.30 & 187.56 & -- & 1.14 &  & W51C & 14.19 & 290.75 & -- & 1.07 \\
 & WComae & 28.23 & 185.38 & -- & 1.20 &  & IC443 & 22.50 & 94.21 & -- & 1.12 \\
 & 1ES1215+303 & 30.10 & 184.45 & -- & 1.26 & Sey2 & ESO139-G12 & -59.94 & 264.41 & -- & 0.82 \\
 & 1ES1218+304 & 30.19 & 185.36 & -- & 1.21 &  & CentaurusA & -43.02 & 201.36 & -- & 0.62 \\
 & Markarian421 & 38.19 & 166.08 & -- & 1.59 & UNID & HESSJ1507-622 & -62.34 & 226.72 & -- & 0.62 \\
Binary & CirX-1 & -57.17 & 230.17 & -- & 0.84 &  & HESSJ1503-582 & -58.74 & 226.46 & -- & 0.62 \\
 & GX339-4 & -48.79 & 255.70 & -- & 0.63 &  & HESSJ1023-575 & -57.76 & 155.83 & 1.5 & 1.08 \\
 & LS5039 & -14.83 & 276.56 & -- & 1.19 &  & HESSJ1614-518 & -51.82 & 243.58 & 0.7 & 0.96 \\
 & SS433 & 4.98 & 287.96 & -- & 0.99 &  & HESSJ1641-463 & -46.30 & 250.26 & -- & 0.78 \\
 & HESSJ0632+057 & 5.81 & 98.24 & 2.7 & 2.40 &  & HESSJ1741-302 & -30.20 & 265.25 & 0.6 & 1.29 \\
FSRQ & S30218+35 & 35.94 & 35.27 & 0.7 & 2.15 &  & HESSJ1826-130 & -13.01 & 276.51 & -- & 1.07 \\
 & B32247+381 & 38.43 & 342.53 & -- & 1.54 &  & HESSJ1813-126 & -12.68 & 273.34 & -- & 0.90 \\
GC & GalacticCentre & -29.01 & 266.42 & 1.1 & 1.36 &  & HESSJ1828-099 & -9.99 & 277.24 & 0.7 & 1.45 \\
PWN & HESSJ1356-645 & -64.50 & 209.00 & 0.4 & 0.98 &  & HESSJ1834-087 & -8.76 & 278.69 & -- & 0.92 \\
 & HESSJ1303-631 & -63.20 & 195.75 & -- & 0.64 &  & 2HWCJ1309-054 & -5.49 & 197.31 & -- & 0.92 \\
 & HESSJ1458-608 & -60.88 & 224.54 & 1.2 & 1.05 &  & 2HWCJ1852+013* & 1.38 & 283.01 & -- & 0.97 \\
 & HESSJ1616-508 & -50.97 & 243.97 & 0.5 & 0.96 &  & 2HWCJ1902+048* & 4.86 & 285.51 & -- & 0.99 \\
 & HESSJ1632-478 & -47.82 & 248.04 & -- & 0.73 &  & MGROJ1908+06 & 6.27 & 286.99 & -- & 1.22 \\
 & VelaX & -45.60 & 128.75 & -- & 0.62 &  & 2HWCJ1829+070 & 7.03 & 277.34 & -- & 1.01 \\
 & HESSJ1831-098 & -9.90 & 277.85 & -- & 0.95 &  & 2HWCJ1907+084* & 8.50 & 286.79 & -- & 1.02 \\
 & HESSJ1837-069 & -6.95 & 279.41 & -- & 1.30 &  & ICPeV & 11.42 & 110.63 & -- & 1.03 \\
 & MGROJ2019+37 & 36.83 & 304.64 & 0.4 & 2.08 &  & 2HWCJ1914+117 & 11.72 & 288.68 & -- & 1.16 \\
Pulsar & PSRB1259-63 & -63.83 & 195.70 & -- & 0.64 &  & 2HWCJ1921+131 & 13.13 & 290.30 & -- & 1.05 \\
 & Terzan5 & -24.90 & 266.95 & -- & 1.09 &  & 2HWCJ0700+143 & 14.32 & 105.12 & -- & 1.48 \\
 & Geminga & 17.77 & 98.47 & 0.9 & 1.75 &  & VERJ0648+152 & 15.27 & 102.20 & -- & 1.57 \\
 & Crab & 22.01 & 83.63 & 0.1 & 1.64 &  & 2HWCJ0819+157 & 15.79 & 124.98 & -- & 1.06 \\
Quasar & PKS1424-418 & -42.10 & 216.98 & 1.1 & 1.04 &  & 2HWCJ1928+177 & 17.78 & 292.15 & -- & 1.26 \\
 & SwiftJ1656.3-3302 & -33.04 & 254.07 & -- & 1.10 &  & 2HWCJ1938+238 & 23.81 & 294.74 & -- & 1.24 \\
 & PKS1622-297 & -29.90 & 246.50 & -- & 0.80 &  & 2HWCJ1949+244 & 24.46 & 297.42 & -- & 1.60 \\
 & PKS0454-234 & -23.43 & 74.27 & -- & 0.84 &  & 2HWCJ1955+285 & 28.59 & 298.83 & -- & 1.18 \\
 & PKS1830-211 & -21.07 & 278.42 & -- & 0.86 &  & 2HWCJ1953+294 & 29.48 & 298.26 & -- & 1.20 \\
 & QSO1730-130 & -13.10 & 263.30 & -- & 0.94 &  & 2HWCJ1040+308 & 30.87 & 160.22 & -- & 1.42 \\
 & PKS0727-11 & -11.70 & 112.58 & 1.3 & 1.59 &  & 2HWCJ2006+341 & 34.18 & 301.55 & -- & 1.38 \\

        \end{tabular}
      }
      \end{minipage}
      
    }
  
\end{table*}

\begin{table}
    \centering
    \caption{The 13 IceCube muon track candidates from the IceCube HESE sample \cite{IC3years, IC4yproc} that are in the field of view of the ANTARES detector. The table gives the equatorial coordinates, the angular error estimate $\beta_\mathrm{IC}$ of the event and the $90\,\%$ C.L. upper limits on flux $\varPhi_0^{90\,\%}$ (in units of $10^{-8}$ GeV cm$^{-2}$ s$^{-1}$).\medskip}
    
    \label{tab:LimitsHESE}
    \begin{tabular}{crrcc}
      HESE ID  &$\delta [\si{\degree}]$     & $\alpha [\si{\degree}]$ & $\beta_\mathrm{IC} [\si{\degree}]$  &  $\varPhi_0^{90\,\%}$  \\
        \hline   
                  3     &        -31.2       &    127.9         &                          1.4             &           2.1          \\
              5     &       -0.4               &          110.6        &                   1.2             &           1.5            \\
              8     &      -21.2               &        182.4          &                   1.3             &           1.7             \\
             13     &       40.3              &          67.9          &                   1.2             &           2.4              \\
             18     &       -24.8            &         345.6            &                   1.3             &           2.0          \\
             23     &     -13.2                 &       208.7         &                   1.9             &           1.7              \\
             28     &        -71.5          &           164.8          &                   1.3             &           1.2             \\
             37     &        20.7           &          167.3           &                   1.2             &           1.7             \\
			 38     &         14.0            &          93.3          &                   1.2             &           2.1              \\
             43     &         -22.0           &       206.6          &                   1.3             &           1.3             \\
             44     &           0.0           &      336.7             &                   1.2             &           1.8          \\
             45     &       -86.3          &      219.0              &                   1.2             &           1.2          \\
             53     &          -37.7              &     239.0         &                   1.2             &           1.6           \\
\end{tabular}%
\end{table}

\subsection{Galactic Center region}
The restricted search region is defined as an ellipse around the Galactic Center with semi-axes of \unit{20}{\degree} in galactic longitude and \unit{15}{\degree} in galactic latitude.
Due to the smaller search area, the search for astrophysical sources is more sensitive than a full sky search because it is less probable for background events to randomly cluster together, mimicking the signature of a signal.
Assuming the usual $E^{-2}$ spectrum, the most significant cluster found in this restricted region is located at $(\alpha,\delta) = (\unit{257.4}{\degree},\unit{-41.0 }{\degree})$ with a pre-trial p-value of 0.09\% and a fitted number of signal events of 2.3. The post-trial significance of this cluster, calculated as in the full sky search but in the restricted region around the Galactic Center, is 60\%. Other spectral indices ($\gamma$ = 2.1, 2.3, 2.5) and source extensions ($\sigma = \unit{0.5}{\degree}$, $\unit{1.0}{\degree}$, $\unit{2.0 }{\degree}$) are considered, yielding different most significant clusters. The source extension is quantified by the $\sigma$ of the gaussian distribution.
For a spectral index of $\gamma$ = 2.5 and a point-source, the most significant cluster is found at $(\alpha,\delta) = (\unit{273.0}{\degree},\unit{-42.2 }{\degree})$, with a pre-trial p-value of 0.02\% and a post-trial significance of 30\%. The distribution of events for these two clusters is shown in Fig.~\ref{fig:AllClusters}-middle-right and bottom-left. The positions of the most significant clusters found for the remaining spectral indices and source extensions considered are within 1$^\circ$ from the position of the latter.

\begin{figure}
     \includegraphics[width=0.48\textwidth]{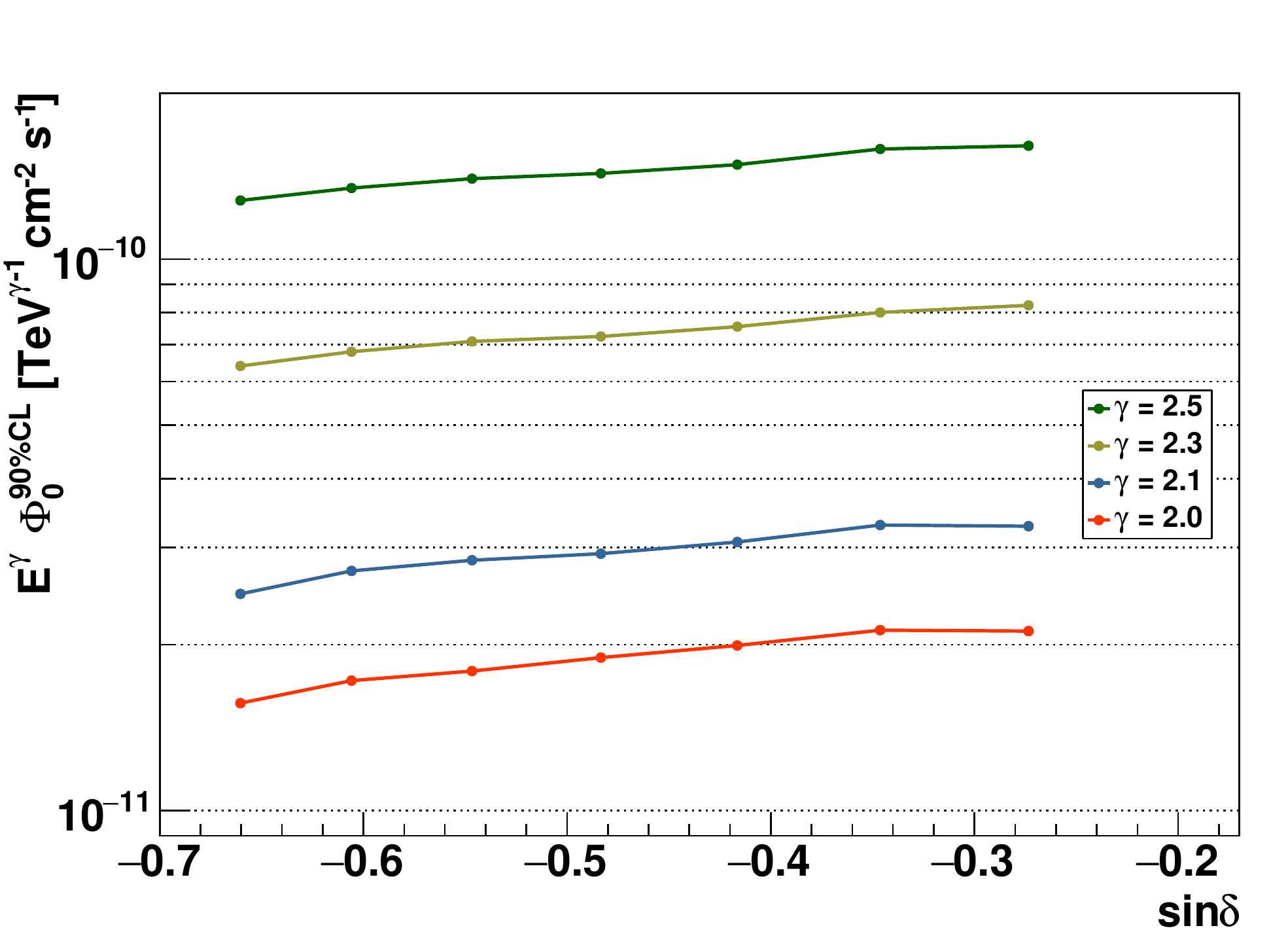}
     \includegraphics[width=0.48\textwidth]{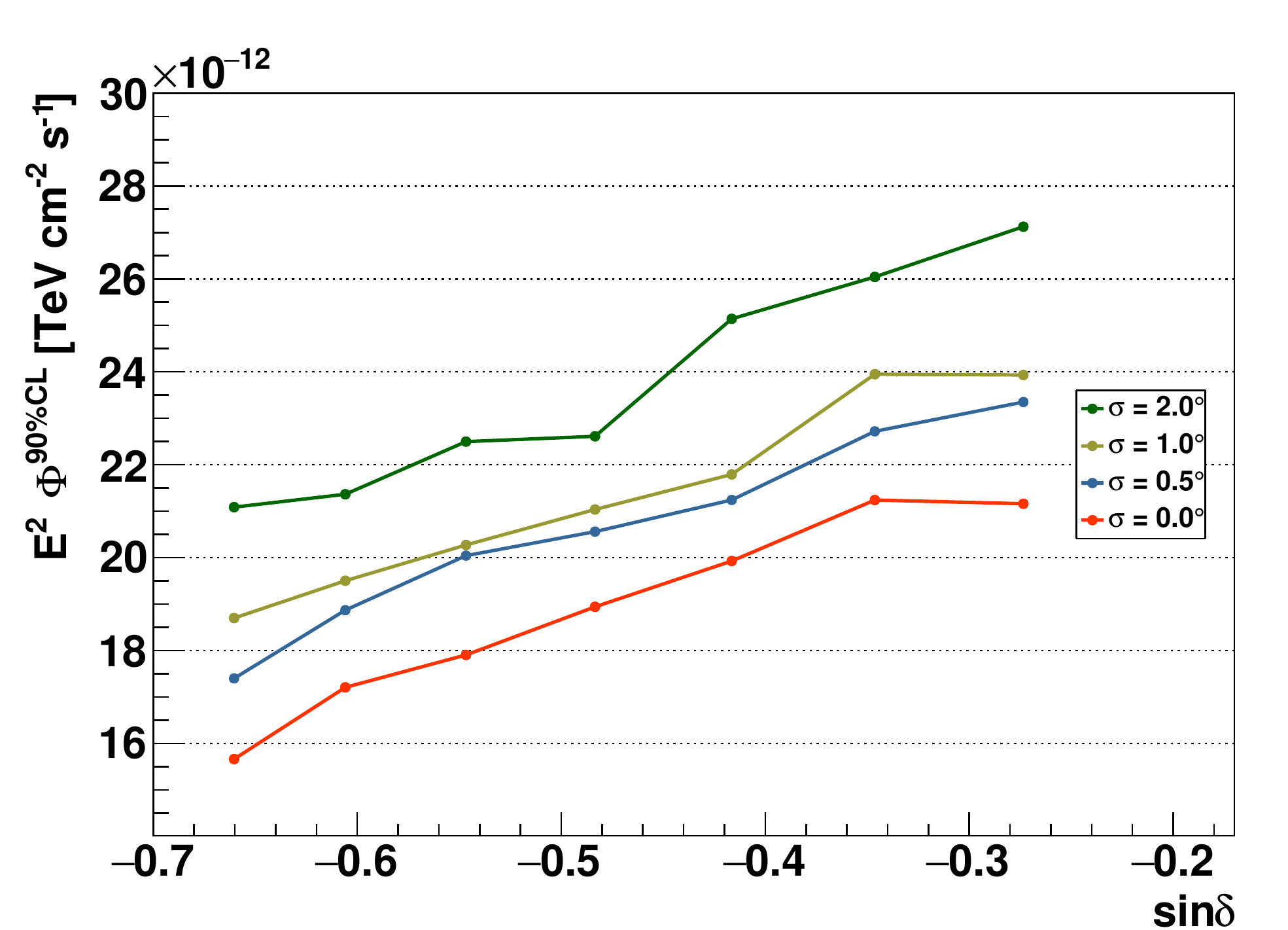}
    \caption{90\% C.L. upper limits of a search restricted to the region around the origin of the galactic coordinates at ($\alpha$, $\delta$) = (266.40$^\circ$,--28.94$^\circ$) assuming different spectral indices for the neutrino flux (left) and different source extensions for $\gamma$ = 2 (right). }
    \label{fig:SensGal}
\end{figure}

The declination-dependent limit of such a restricted point-like source search is shown in Fig.~\ref{fig:SensGal}, both for different energy spectral indices $\gamma$ and different source extensions. The upper limits increase with increasing values of $\gamma$ and with the source extension. A softer energy spectrum of cosmic neutrinos (larger values of the spectral index $\gamma$) is less distinguishable from the spectrum of atmospheric neutrinos, as is a source with a larger extension.
For a softer spectrum, fewer neutrinos are emitted by the source within an energy range in which they can be statistically separated from atmospheric neutrinos. The flux required at the normalization point for a significant detection is therefore larger.

\subsection{Sagittarius A*}
\begin{figure}[!h]
    \centering
    \includegraphics[width=0.6\textwidth]{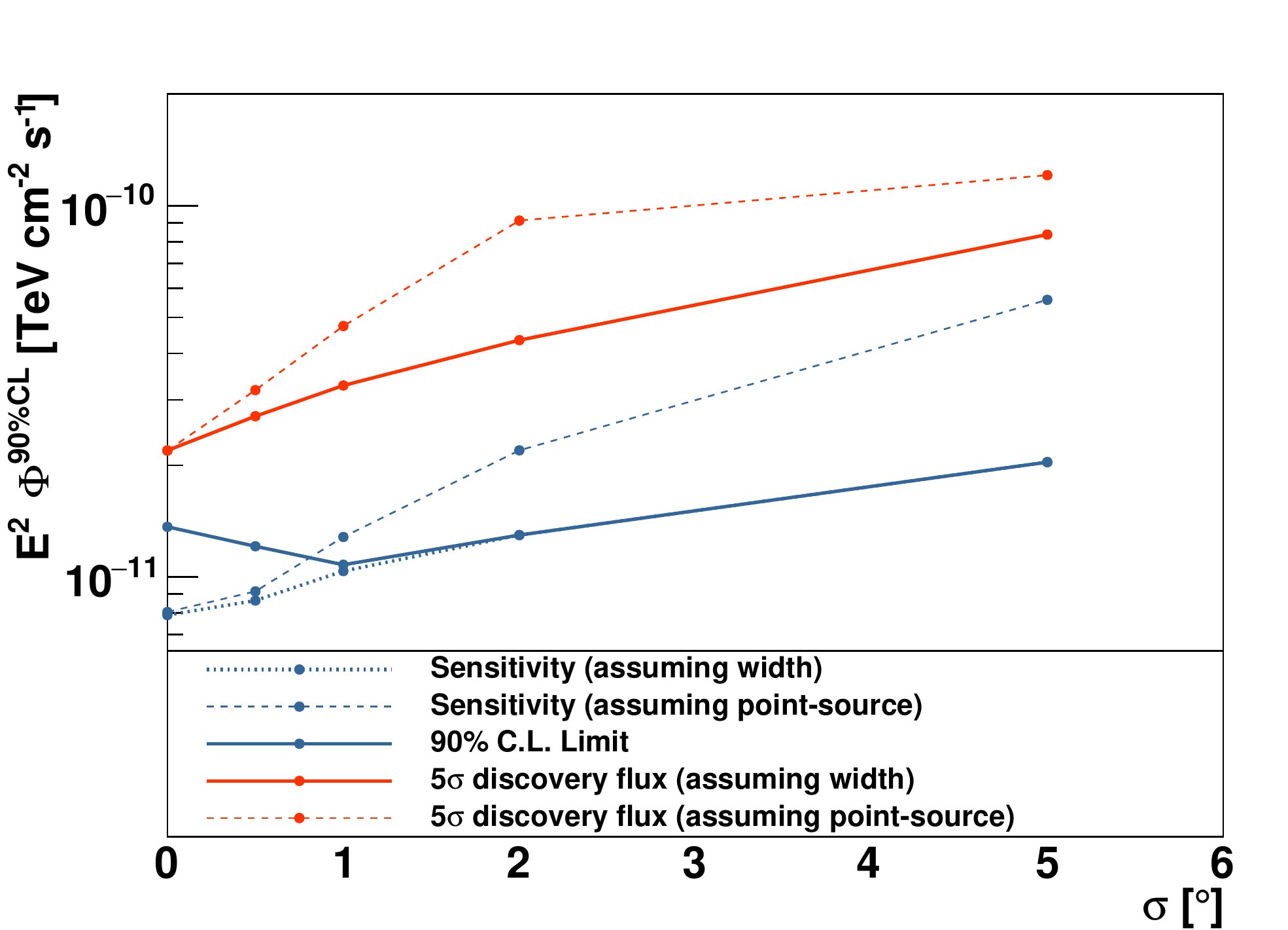}
    \caption{Discovery flux (dotted red), median sensitivity (dotted blue) and 90\,\% C.L. upper limits (green) for a search for an extended source at Sagittarius A* at $(\alpha,\delta) = (\unit{266.42}{\degree},\unit{-29.01}{\degree}$) assuming different angular extensions $\sigma$. The dashed lines correspond to the point-like source assumption.
    }
    \label{fig:SensExt}
\end{figure}

Super-massive black holes are strong candidates to be accelerators of very-high energy cosmic rays and therefore for cosmic neutrino production \cite{LightHouseSgrA}.
Additionally, due to the high concentration of candidate sources and gas around the Galactic Center (GC), it is probable that an extended signal from that region will be detected before identifying individual point-like sources.
For this reason, Sagittarius A*, located at $(\alpha,\delta) = (\unit{266.42}{\degree},\unit{-29.01}{\degree}$), is investigated as an extended source with widths between \unit{0.5}{\degree} and \unit{5}{\degree}. The cluster of events around Sagittarius A* reconstructed by ANTARES is shown in Fig.~\ref{fig:AllClusters}-bottom-right.
The sensitivity and upper limits for the assumption of different source extensions can be seen in Fig.~\ref{fig:SensExt}. The sensitivity degrades with increasing extension but an improvement of up to a factor of $2.7$ can be achieved by assuming an extended source with the simulated extension. The largest excess above the background is found at an extension of \unit{0}{\degree} with a pre-trial p-value of 22\%. 

\section{Systematic uncertainties}
\label{sec:SYST}

The effects of systematic uncertainties on the absolute pointing accuracy, angular resolution, acceptance and the background rate distribution of events are evaluated.

\emph{Absolute Pointing Accuracy Uncertainty.} An uncertainty of 0.13$^\circ$ and 0.06$^\circ$ on the horizontal ($\phi$) and vertical ($\theta$) directions, respectively, was established in a previous study \cite{PointingAcc}. To take this into account, randomly generated offsets have been added to the $\phi$ and $\theta$ variables of the simulated events. The offsets are generated according to two Gaussian distributions with the aforementioned uncertainties as sigmas.

\emph{Angular Resolution Uncertainty.} The angular resolution of the track reconstruction algorithm can be affected by the accuracy of the detected hit times. A smearing of these times was performed in simulations leading to a 15\% degradation on the angular resolution in the track channel \cite{antPS4y}. For neutrinos of the shower sample, the reconstruction of the direction depends most significantly on the recorded charge. A smearing %of 30\% (50\%) 
in the measured charges \cite{ChargeResol} leads to a 12\% degradation of the angular resolution for the shower channel.

\emph{Acceptance Uncertainty.} A 15\% uncertainty on the acceptance has been considered for the calculation of the reported fluxes. This uncertainty was calculated after performing simulations with a reduction of the OM efficiency by 15\% \cite{antPS4y}.

\emph{Background Uncertainty.} In order to account for possible systematic uncertainties on the background, the distribution of the background rates in Fig.~\ref{fig:background} are parametrized by two different spline functions, $R(\delta)$ and $B(\delta)$ (the red and blue lines). The declination-dependent distribution of background events of the pseudo-experiments is determined as $\mathscr B(\delta) = B(\delta) + r \cdot ( R(\delta) - B(\delta) )$, with $r$ being a random number drawn for each pseudo-experiment from a uniform distribution between -1 and 1.

It is found that not considering these uncertainties would improve the median sensitivity at $90\,\%$ C.L. and the $5 \sigma$ discovery potential by less than $5\,\%$. 

\section{Conclusion and outlook}\label{sec:CONCL}
Various searches for cosmic neutrino sources using combined information from the track and shower channels have been presented. 
These searches provide the most sensitive limits for a large fraction of the Southern Sky, especially at neutrino energies below 100 TeV. No significant evidence of cosmic neutrino sources has been found. The IceCube HESE accumulation reported near the Galactic Center could neither be totally attributed to a point-like source nor to an extended source.

The most significant cluster in the full sky search is located at $(\alpha,\delta) = (\unit{343.8}{\degree}, \unit{23.5}{\degree})$ with a post-trial significance of $5.9\,\%$ or $1.9\sigma$. 

Upper limits on the neutrino flux from 106 astrophysical candidates and 13 IceCube muon tracks have been presented.
The most significant source candidate is HESSJ0632+057 -- located at $(\alpha,\delta) = (\unit{98.24}{\degree},\unit{5.81}{\degree})$ -- with a post-trial significance of $1.5\sigma$. The upper limit on the signal from this candidate is $E^2 d\varPhi/dE = \unit{2.40e-8}{\unitFluxNorm}$.
 
The most significant cluster of events close to the Galactic Center when assuming a point-like source with an $E^{-2}$ energy spectrum is located at $(\alpha,\delta) = (\unit{-102.6}{\degree},\unit{-41.0 }{\degree})$ with a post-trial p-value of 60\%. 

Sagittarius A* as a possible extended source has been investigated. Upper limits for the flux and number of events assuming a Gaussian morphology with different extensions have been presented. The largest excess over the background is observed at an angular extension of \unit{0}{\degree} with a pre-trial p-value value of 22\%.\\

The KM3NeT/ARCA neutrino telescope \cite{km3netLOI}, which is currently under construction, will combine a cubic kilometer-sized detector with the same high visibility towards the Galactic Center as ANTARES. It is expected that this detector will be able to make definite statements about a neutrino flux from several Galactic candidates within a few years of operation.

\section*{Acknowledgements}

The authors acknowledge the financial support of the funding agencies:
% France:
Centre National de la Recherche Scientifique (CNRS), Commissariat \`a
l'\'ener\-gie atomique et aux \'energies alternatives (CEA),
Commission Europ\'eenne (FEDER fund and Marie Curie Program),
Institut Universitaire de France (IUF), IdEx program and UnivEarthS
Labex program at Sorbonne Paris Cit\'e (ANR-10-LABX-0023 and
ANR-11-IDEX-0005-02), Labex OCEVU (ANR-11-LABX-0060) and the
A*MIDEX project (ANR-11-IDEX-0001-02),
R\'egion \^Ile-de-France (DIM-ACAV), R\'egion
Alsace (contrat CPER), R\'egion Provence-Alpes-C\^ote d'Azur,
D\'e\-par\-tement du Var and Ville de La
Seyne-sur-Mer, France;
% Germany: 
Bundesministerium f\"ur Bildung und Forschung
(BMBF), Germany; 
% Italy
Istituto Nazionale di Fisica Nucleare (INFN), Italy;
% Netherlands
Stichting voor Fundamenteel Onderzoek der Materie (FOM), Nederlandse
organisatie voor Wetenschappelijk Onderzoek (NWO), the Netherlands;
% Russia
Council of the President of the Russian Federation for young
scientists and leading scientific schools supporting grants, Russia;
% Romania
National Authority for Scientific Research (ANCS), Romania;
% Spain 
Mi\-nis\-te\-rio de Econom\'{\i}a y Competitividad (MINECO):
Plan Estatal de Investigaci\'{o}n (refs. FPA2015-65150-C3-1-P, -2-P and -3-P, (MINECO/FEDER)), Severo Ochoa Centre of Excellence and MultiDark Consolider (MINECO), and Prometeo and Grisol\'{i}a programs (Generalitat
Valenciana), Spain; 
% Marocco
Ministry of Higher Education, Scientific Research and Professional Training, Morocco.
% A.O.B.:
We also acknowledge the technical support of Ifremer, AIM and Foselev Marine
for the sea operation and the CC-IN2P3 for the computing facilities.

\clearpage

\appendix

\section{Shower selection cut parameters}
\label{appendix}

\begin{figure}[t]
 
        \includegraphics[width=0.48\textwidth]{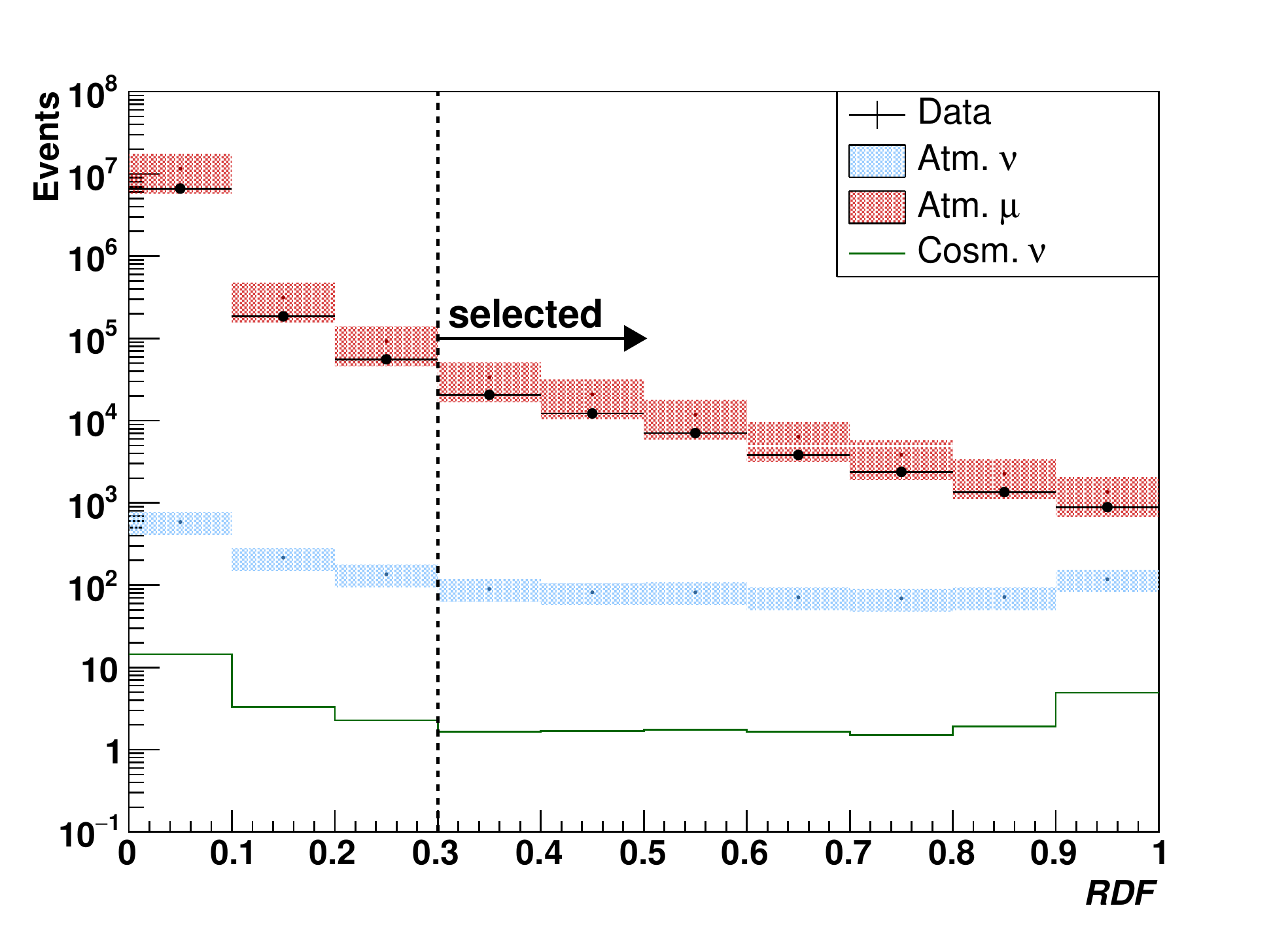}
        \includegraphics[width=0.48\textwidth]{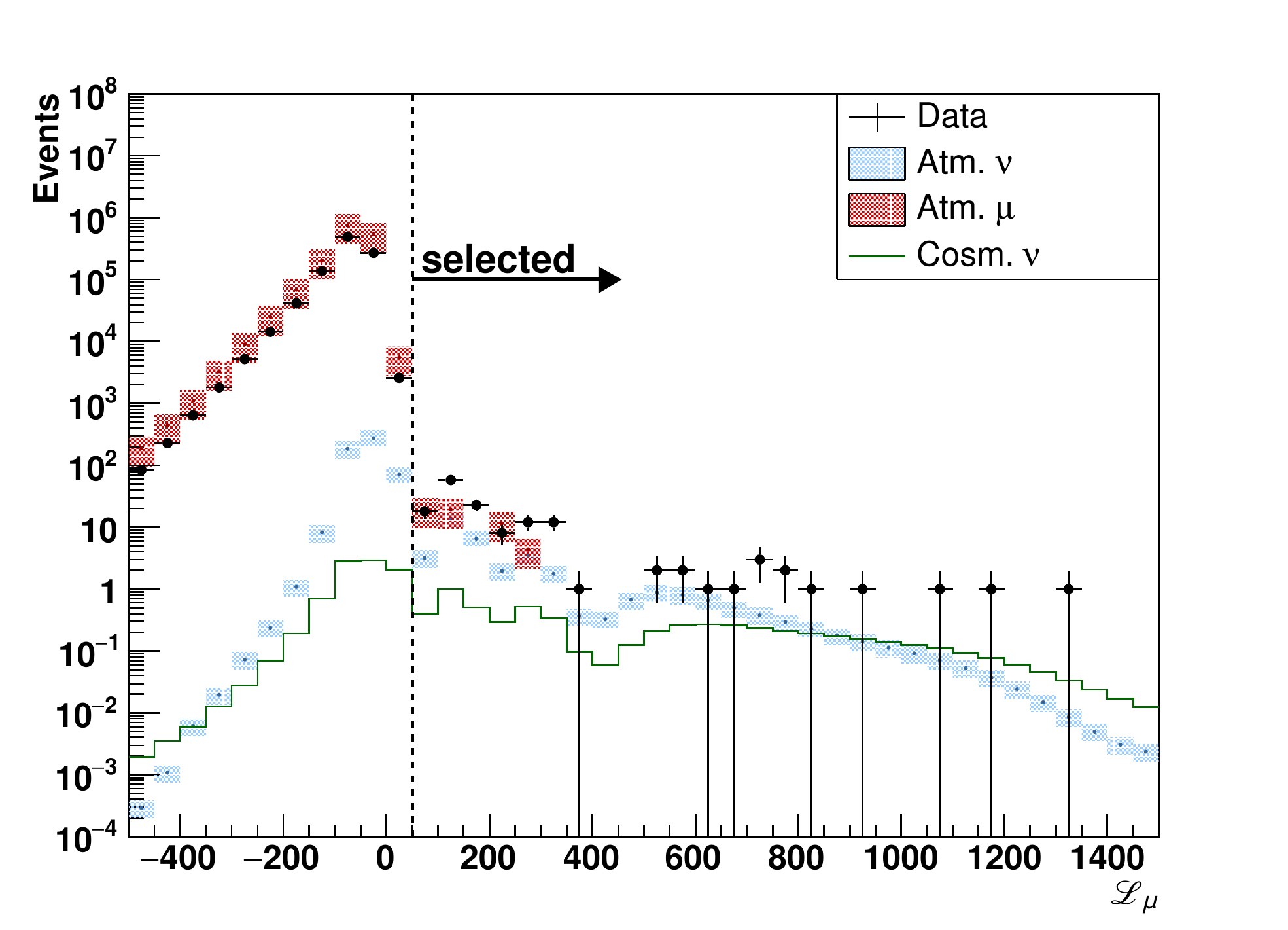}

    \caption{$RDF$ parameter for data, cosmic neutrinos and atmospheric background (left). This figure corresponds to the event distributions after all the cuts prior to the $RDF$ listed in Table~\ref{tab:ShowSel}. Right: muon likelihood ratio parameter for data, cosmic neutrinos and atmospheric background. This figure corresponds to the event distributions after the $RDF$ and all previous cuts listed in Table~\ref{tab:ShowSel}. In both figures the dashed vertical line indicates the cut value. }
    \label{fig:shower_cuts}
\end{figure}

Additional details concerning some of the parameters referred to Table~\ref{tab:ShowSel} used to define selection criteria for the shower channel events are given in this section.

\begin{description}

    \item [Interaction vertex.] 
Reconstructing \hskip .2em atmospheric \hskip .2em muons \hskip .2em with \hskip .2em a \hskip .2em shower algorithm often results in reconstructed vertex positions that lie far away from the instrumented volume of the detector. This can be approximated with a cylindrical structure with a height of 350 m and a radius of 180 m.  
A cut on the radial distance of the reconstructed shower position from the vertical axis of the detector ($R_{sh} <$ 300 m) and on the vertical distance above the center of the detector ($| Z_{sh }| < $ 250 m ) are applied. 
        \item [RDF.] 
        A different shower reconstruction algorithm was originally developed for diffuse flux analyses \cite{Dusj-Paper}. Among all available quality parameters provided by this reconstruction chain, a subset of five parameters that showed a high potential to separate atmospheric muon tracks from shower events was chosen as input in a Random Decision Forest ($RDF$) classification. 
        The distribution of the $RDF$ parameter for cosmic neutrinos and atmospheric muons and neutrinos after applying the cuts prior to the RDF cut is shown in Fig.~\ref{fig:shower_cuts}-left. 
        Only shower events with $RDF > 0.3$ are used in this analysis.

    \item [Muon likelihood.] An additional likelihood function has been developed to discriminate between neutrinos that produce showers and the background of atmospheric muons.
    This likelihood considers only hits that coincide with another hit on the same storey (which contains a triplet of PMTs at the same position in the detector line) within \unit{20}{\nano\second}. Its probability density function is based on the time residual (time difference between the detected and the expected hit from the assumption of a point-like light emission from the simulated vertex position without photon scattering) $t_\mathrm{res}$ of the hits, the number $N$ of on-time hits ($-20$ ns $ < t_\mathrm{res}  < 60$ ns) and the distance $d$ of the hits to the reconstructed shower position.
    The parameter to distinguish between showers and muons then is
    $$ \mathscr L_\mathrm{\mu} = \sum_\mathrm{Hits}\log\{P_\mathrm{sig} / P_\mathrm{bkg}\},$$
    with $P_\mathrm{sig}\ = P(N, d, t_\mathrm{res} | \mathrm{\nu})$ and $P_\mathrm{bkg} = P(N, d, t_\mathrm{res} | \mathrm{\mu})$.\\
    The distribution for this quantity plotted for atmospheric muons and cosmic showers after all cuts prior to the muon likelihood cut is shown in Fig.~\ref{fig:shower_cuts}-right. Shower events with $\mathscr L_\mathrm{\mu} < 50$ are excluded from the analysis. This method further reduces the number of atmospheric muons by more than two orders of magnitude. 
\end{description}%

\bibliographystyle{utphys}
\bibliography{Bibliography}

\end{document}